\documentclass[12pt]{article}
\usepackage{amsmath,amssymb,amsthm,amsxtra,overpic,bbm,bm,epsfig,subfigure}
\usepackage{hyperref}
\usepackage{mathrsfs}
\usepackage{graphicx}
\usepackage{multirow}
\usepackage{color}
\usepackage{comment}
\usepackage{epstopdf}
\usepackage{float}
\numberwithin{equation}{section}
\usepackage{cite}
\usepackage{hyperref}
\usepackage{url}
\usepackage{slashed,stmaryrd}

\def\thefootnote{\fnsymbol{footnote}}

\begin{document}

\vspace{0.2cm}
\begin{center}
{\Large\bf Sufficient and Necessary Conditions for CP Conservation in the Case of Degenerate Majorana Neutrino Masses}
\end{center}
\vspace{0.2cm}

\begin{center}
{\bf Bingrong Yu~$^{a,~b}$}~\footnote{E-mail: yubr@ihep.ac.cn},
\quad
{\bf Shun Zhou~$^{a,~b}$}~\footnote{E-mail: zhoush@ihep.ac.cn (corresponding author)}
\\
\vspace{0.2cm}
{\small $^a$Institute of High Energy Physics, Chinese Academy of
Sciences, Beijing 100049, China \\
$^b$School of Physical Sciences, University of Chinese Academy of Sciences, Beijing 100049, China}
\end{center}

\vspace{1.5cm}

\begin{abstract}
In this paper, we carry out a systematic study of the sufficient and necessary conditions for CP conservation in the leptonic sector with massive Majorana neutrinos. In particular, the emphasis is placed on the number of CP-violating phases in the presence of a partial mass degeneracy (e.g., $m^{}_1 = m^{}_2 \neq m^{}_3$) or a complete mass degeneracy $m^{}_1 = m^{}_2 = m^{}_3$, where $m^{}_i$ (for $i = 1, 2, 3$) stand for the masses of three ordinary neutrinos. In the canonical seesaw model with three right-handed neutrino singlets, CP-violating phases in the special case of a partial (e.g., $M^{}_1 = M^{}_2 \neq M^{}_3$) or complete (i.e., $M^{}_1 = M^{}_2 = M^{}_3$) mass degeneracy of three heavy Majorana neutrinos are also examined. In addition, we derive the renormalization-group equations of the weak-basis invariants in the effective theory with a general mass spectrum of Majorana neutrinos, to which the solutions establish the direct connection between CP violation at low- and high-energy scales.
\end{abstract}

\newpage

\def\thefootnote{\arabic{footnote}}
\setcounter{footnote}{0}

\section{Introduction}

Neutrino oscillation experiments have firmly established that neutrinos are indeed massive and lepton flavors are significantly mixed~\cite{Tanabashi:2018oca, Xing:2019vks}. One main goal of future long-baseline accelerator neutrino oscillation experiments is to discover CP violation in the leptonic sector and precisely measure the relevant CP-violating phase~\cite{Branco:2011zb}. In order to account for tiny neutrino masses, one can go beyond the standard model (SM) by introducing three right-handed neutrino singlets $N^{}_{i {\rm R}}$ (for $i = 1, 2, 3$). Then the ${\rm SU}(2)^{}_{\rm L}\times {\rm U}(1)^{}_{\rm Y}$ gauge-invariant Lagrangian for lepton masses, flavor mixing and CP violation can be written as
\begin{eqnarray}\label{eq:seesawlag}
{\cal L}^{}_{\rm lepton} = -\overline{\ell^{}_{\rm L}} Y^{}_l l^{}_{\rm R} H - \overline{\ell^{}_{\rm L}} Y^{}_\nu \tilde{H} N^{}_{\rm R} - \frac{1}{2} \overline{N^{\rm C}_{\rm R}} M^{}_{\rm R} N^{}_{\rm R} + {\rm h.c.} \; ,
\end{eqnarray}
where $\ell^{}_{\rm L} \equiv (\nu^{}_{\rm L}, l^{}_{\rm L})^{\rm T}$ and $\tilde{H} \equiv {\rm i}\sigma^{}_2 H^*$ with $H\equiv (\varphi^+, \varphi^0)^{\rm T}$ are the left-handed lepton doublet and the Higgs doublet, $Y^{}_l$ and $Y^{}_\nu$ are the charged-lepton and Dirac neutrino Yukawa coupling matrices, and $M^{}_{\rm R}$ is the Majorana mass matrix for right-handed neutrino singlets. Note that $N^{\rm C}_{\rm R} \equiv {\cal C}\overline{N^{}_{\rm R}}^{\rm T}$ has been defined with ${\cal C} \equiv {\rm i}\gamma^2 \gamma^0$ being the charge-conjugation matrix. As the Higgs field acquires its vacuum expectation value $\langle \varphi^0 \rangle = v/\sqrt{2}$ with $v \approx 246~{\rm GeV}$ and the gauge symmetry is spontaneously broken down, the charged-lepton mass matrix and the Dirac neutrino mass matrix are then given by $M^{}_l \equiv Y^{}_l v/\sqrt{2}$ and $M^{}_{\rm D} \equiv Y^{}_\nu v/\sqrt{2}$, respectively.

In such a minimal extension of the SM, three ordinary neutrinos are massive Majorana particles, namely, they are their own antiparticles~\cite{Majorana:1937vz, Racah:1937qq}. The lepton mass spectrum, flavor mixing and CP violation at the low-energy scale are then governed by the following effective Lagrangian
\begin{eqnarray}\label{eq:effectivelag}
{\cal L}^\prime_{\rm lepton} = - \overline{l^{}_{\rm L}} M^{}_l l^{}_{\rm R} - \frac{1}{2} \overline{\nu^{}_{\rm L}} M^{}_\nu \nu^{\rm C}_{\rm L} + \frac{g}{\sqrt{2}} \overline{l^{}_{\rm L}} \gamma^\mu \nu^{}_{\rm L} W^-_\mu + {\rm h.c.} \; ,
\end{eqnarray}
where the effective mass matrix for three light Majorana neutrinos is given by the famous seesaw formula $M^{}_\nu = - M^{}_{\rm D} M^{-1}_{\rm R} M^{\rm T}_{\rm D}$~\cite{Minkowski, Yanagida, Gell-Mann, Glashow, Mohapatra}, which is in general complex and symmetric, and the last term stands for the charged-current weak interaction with $g$ being the gauge coupling constant of the ${\rm SU}(2)^{}_{\rm L}$ gauge group. As the Majorana mass term of right-handed neutrinos is not subject to the spontaneous gauge symmetry breaking, the smallness of light Majorana neutrino masses ${\cal O}(M^{}_\nu) \lesssim 0.1~{\rm eV}$ can be ascribed to the largeness of heavy Majorana neutrino masses ${\cal O}(M^{}_{\rm R}) \gtrsim 10^{14}~{\rm GeV}$ with ${\cal O}(M^{}_{\rm D}) \sim 10^{2}~{\rm GeV}$. After diagonalizing the lepton mass matrices via $V^\dagger_l M^{}_l V^\prime_l = \widehat{M}^{}_l \equiv {\rm diag}\{m^{}_e, m^{}_\mu, m^{}_\tau\}$ and $V^\dagger_\nu M^{}_\nu V^*_\nu = \widehat{M}^{}_\nu \equiv {\rm diag}\{m^{}_1, m^{}_2, m^{}_3\}$, where $V^{}_l$, $V^\prime_l$ and $V^{}_\nu$ are $3\times 3$ unitary matrices, and converting into the mass basis, we can obtain the leptonic flavor mixing matrix or the Pontecorvo-Maki-Nakagawa-Sakata (PMNS) matrix $U = V^\dagger_l V^{}_\nu$~\cite{Pontecorvo:1957cp, Maki:1962mu}, which then appears in the charged-current interaction as the origin of lepton flavor mixing and CP violation.

Since the discovery of leptonic CP violation is the primary goal of future neutrino oscillation experiments and it may also be connected to cosmological matter-antimatter asymmetry~\cite{Sakharov:1967dj, Fukugita:1986hr, Davidson:2008bu, Bodeker:2020ghk}, it is interesting to establish the sufficient and necessary conditions for CP conservation in the leptonic sector with massive Majorana neutrinos. Only when these conditions are spoiled in a specific model of neutrino masses can one explain the leptonic CP violation and associate it with the dynamical generation of cosmological matter-antimatter asymmetry. This task has already been taken up in the literature, particularly by Branco and his collaborators~\cite{Branco:1986gr} in the language of the so-called weak-basis (WB) invariants~\cite{Bernabeu:1986fc}. The leptonic CP violation in terms of WB invariants has been discussed first by Pilaftsis in the context of resonant leptogenesis~\cite{Pilaftsis:1997jf} and subsequently by several other authors~\cite{Branco:2001pq,Cirigliano:2006nu}. The central idea of this approach is to define the general CP transformation, which is actually combination of the ordinary CP transformation and the flavor-basis transformation. More explicitly, if the Lagrangian in Eq.~(\ref{eq:effectivelag}) is invariant under the following transformations~\cite{Branco:1986gr}
\begin{eqnarray}\label{eq:transformation}
l^{}_{\rm L} \to  U^{}_{\rm L} {\cal C}l^*_{\rm L} \; , \quad \nu^{}_{\rm L} \to U^{}_{\rm L} {\cal C}\nu^*_{\rm L} \; , \quad
l^{}_{\rm R} \to  U^{}_{\rm R} {\cal C}l^*_{\rm R} \; , \quad
W^-_\mu \to  -(-1)^{\delta^{}_{0\mu}} W^+_\mu \; ,
\end{eqnarray}
where the asterisk ``$*$" indicates the complex conjugation and $\delta^{}_{0\mu}$ (for $\mu = 0, 1, 2, 3$) stands for the Kronecker delta, while $U^{}_{\rm L}$ and $U^{}_{\rm R}$ are two arbitrary $3\times 3$ unitary matrices in the flavor space, then the sufficient and necessary conditions for CP conservation are equivalent to the existence of two unitary matrices $U^{}_{\rm L}$ and $U^{}_{\rm R}$ such that the identities below~\cite{Branco:1986gr}
\begin{eqnarray}\label{eq:necsuff}
U^\dagger_{\rm L} M^{}_\nu U^*_{\rm L} = - M^*_\nu \; , \quad U^\dagger_{\rm L} M^{}_l U^{}_{\rm R} = M^*_l \;,
\end{eqnarray}
are satisfied. With the help of Eq.~(\ref{eq:necsuff}), one can find out the minimal set of sufficient and necessary conditions for CP conservation in the leptonic sector in terms of WB invariants~\cite{Dreiner:2007yz}
\begin{eqnarray}
{\cal I}^{}_1 &\equiv& {\rm Tr}\left\{ \left[H^{}_\nu, H^{}_l \right]^3\right\} = 0 \; , \label{eq:I1} \\
{\cal I}^{}_2 &\equiv& {\rm Im}\left\{{\rm Tr}\left[H^{}_l H^{}_\nu G^{}_{l\nu}\right]\right\} = 0 \; , \label{eq:I2} \\
{\cal I}^{}_3 &\equiv& {\rm Tr}\left\{ \left[G^{}_{l\nu}, H^{}_l \right]^3\right\} = 0 \; , \label{eq:I3}
\end{eqnarray}
where $H^{}_l \equiv M^{}_l M^\dagger_l$, $H^{}_\nu \equiv M^{}_\nu M^\dagger_\nu$ and $G^{}_{l\nu} \equiv M^{}_\nu H^*_l M^\dagger_\nu$ have been introduced.

It has been pointed out in Ref.~\cite{Yu:2019ihs} that those conditions in Eqs.~(\ref{eq:I1})-(\ref{eq:I3}) are not sufficient to guarantee CP conservation in general. A numerical counter example has been given therein to illustrate that CP violation still exists even when all three conditions in Eqs.~(\ref{eq:I1})-(\ref{eq:I3}) are satisfied. For this reason, a new set of three invariants \{${\cal I}_1$, ${\cal I}_2$, ${\cal I}_4$\} has been suggested in Ref.~\cite{Yu:2019ihs} with ${\cal I}_4$ defined as
\begin{eqnarray}
{\cal I}_4 \equiv {\rm Im}\left\{{\rm Tr}\left[H^{}_l H^{2}_\nu G^{}_{l\nu}\right]\right\} = 0 \; , \label{eq:I4}
\end{eqnarray}
which can guarantee CP conservation at least in the experimentally allowed parameter space of lepton masses and mixing angles.\footnote{Here the number $n$ of vanishing WB invariants ${\cal I}^{}_1 = {\cal I}^{}_2 = {\cal I}^{}_4 = 0$ to guarantee CP conservation coincides with the number of CP phases in the theory, i.e., $n = 3$ in the present case. However, this is valid under the assumption that the lepton masses and the flavor mixing angles do not take any contrived values. More explicitly, we assume that all the other physical parameters take the values within their experimentally allowed parameter space. If this is not assumed, as pointed out in Ref.~\cite{Yu:2019ihs}, then there may still exist CP violation in a theory with $n$ CP phases even when the properly chosen $n$ WB invariants vanish.} Note that the invariance under the general CP transformations in Eq.~(\ref{eq:transformation}) requires $U^\dagger_{\rm L} H^{}_l U^{}_{\rm L} = H^*_l$, $U^\dagger_{\rm L} H^{}_\nu U^{}_{\rm L} = H^*_\nu$ and $U^\dagger_{\rm L} G^{}_{l\nu} U^{}_{\rm L} = G^*_{l\nu}$ according to Eq.~(\ref{eq:necsuff}). By using these transformation rules, one can immediately prove that ${\cal I}^{}_i$ (for $i = 1, 2, 3, 4$) are indeed WB invariants. Given the lepton mass matrices $M^{}_l$ and $M^{}_\nu$ in a concrete model, the advantage of these WB invariants is to remove the ambiguity of flavor-basis transformations in ensuring whether CP conservation is present.

In this work, we aim to derive the sufficient and necessary conditions for CP conservation in the leptonic sector, and especially focus on the scenario of a partially- or completely-degenerate neutrino mass spectrum~\cite{Branco:1998bw, Mei:2003gu}. The number of flavor mixing angles and CP-violating phases in these special cases will be clarified. In addition, we investigate the radiative corrections to leptonic CP violation by using the renormalization-group equations of the WB invariants. On the other hand, since neutrino oscillation experiments indicate that any two of three light neutrinos cannot be exactly degenerate in mass, we consider the mass degeneracy for heavy Majorana neutrinos in the canonical seesaw models~\cite{Branco:2001pq, Rebelo:2018qsj} and explore the implications of such a partial or complete mass degeneracy for the leptonic CP violation at low- and high-energy scales.

The remaining part of our paper is structured as follows. In Sec.~\ref{sec:low}, we recall the sufficient and necessary conditions for the CP violation in the low-energy effective theory of lepton masses and flavor mixing, and pay a particular attention to the cases of a partial or complete degeneracy in neutrino masses. The renormalization-group equations of the WB invariants will be derived and used to study the running behaviors of CP phases.  Then, we apply the formalism for light Majorana neutrinos to the case of heavy Majorana neutrinos in Sec. ~\ref{sec:high}. The full set of WB invariants for CP conservation will be given and utilized to analyze the possible connection between CP violation at low- and high-energy scales. Finally, we summarize our main conclusions in Sec.~\ref{sec:summary}.

\section{Low-energy Effective Theory}\label{sec:low}

At the low-energy scale, leptonic CP violation arises from the complex mass matrices of charged leptons and light Majorana neutrinos, as indicated in Eq.~(\ref{eq:effectivelag}). If three Majorana neutrinos are not degenerate in mass, the minimal set of sufficient and necessary conditions for CP conservation have already been given in Eqs.~(\ref{eq:I1}), (\ref{eq:I2}) and (\ref{eq:I4}). Although the details can be found in Ref.~\cite{Yu:2019ihs}, we briefly summarize the key points concerning these sufficient and necessary conditions for CP conservation in the case of nondegenerate neutrino masses in order to establish our notations. As the WB invariants are by definition independent of basis transformations in the flavor space, it should be kept in mind that one can calculate them in any convenient basis and the final results depend only on physical parameters.

In the mass basis of charged leptons and light Majorana neutrinos, the CP-violating phases are contained in the PMNS matrix~\cite{Tanabashi:2018oca}, which is usually parametrized in terms of three mixing angles $\{\theta^{}_{12}, \theta^{}_{13}, \theta^{}_{23}\}$, one Dirac-type CP phase $\delta$ and two Majorana-type CP phases $\{\rho, \sigma\}$, namely,
\begin{eqnarray}\label{eq:parametrization}
U = \left( \begin{matrix} c^{}_{13} c^{}_{12} & c^{}_{13} s^{}_{12} & s^{}_{13} e^{-{\rm i}\delta} \cr -s_{12}^{} c_{23}^{} - c_{12}^{} s_{13}^{} s_{23}^{} e^{{\rm i}\delta}_{} & + c_{12}^{} c_{23}^{} - s_{12}^{} s_{13}^{} s_{23}^{} e^{{\rm i}\delta}_{} & c_{13}^{} s_{23}^{} \cr + s_{12}^{} s_{23}^{} - c_{12}^{} s_{13}^{} c_{23}^{} e^{{\rm i}\delta}_{} & - c_{12}^{} s_{23}^{} - s_{12}^{} s_{13}^{} c_{23}^{} e^{{\rm i}\delta}_{} & c_{13}^{} c_{23}^{} \end{matrix} \right) \cdot \left(\begin{matrix} e^{{\rm i}\rho} & 0 & 0 \cr 0 & e^{{\rm i}\sigma} & 0 \cr 0 & 0 & 1\end{matrix}\right) \; ,
\end{eqnarray}
where $c^{}_{ij} \equiv \cos \theta^{}_{ij}$ and $s^{}_{ij} \equiv \sin \theta^{}_{ij}$ (for $ij = 12, 13, 23$) have been defined. Therefore, the invariants $\{{\cal I}^{}_1, {\cal I}^{}_2, {\cal I}^{}_4\}$ can be expressed in terms of the charged-lepton masses $\{m^{}_e, m^{}_\mu, m^{}_\tau\}$, neutrino masses $\{m^{}_1, m^{}_2, m^{}_3\}$, leptonic flavor mixing angles $\{\theta^{}_{12}, \theta^{}_{13}, \theta^{}_{23}\}$ and CP-violating phases $\{\delta, \rho, \sigma\}$. It is straightforward to verify that the invariant ${\cal I}^{}_1$ in Eq.~(\ref{eq:I1}) can be written as
\begin{eqnarray}
{\cal I}_1 = - 6 {\rm i} \Delta^{}_{21} \Delta^{}_{31} \Delta^{}_{32} \Delta^{}_{e\mu} \Delta^{}_{\mu\tau} \Delta^{}_{\tau e} {\cal J} \; , \label{eq:I1expr}
\end{eqnarray}
where $\Delta^{}_{ij} \equiv m^2_i - m^2_j$ (for $i, j = 1, 2, 3$) are neutrino mass-squared differences, $\Delta^{}_{\alpha \beta} \equiv m^2_\alpha - m^2_\beta$ (for $\alpha, \beta = e, \mu, \tau$) are charged-lepton mass-squared differences, and ${\cal J} \equiv {\rm Im}\left[U^{}_{e1} U^*_{e2} U^*_{\mu 1} U^{}_{\mu 2}\right]$ is the Jarlskog invariant for CP violation in leptonic sector~\cite{Jarlskog:1985ht, Wu:1985ea, Cheng:1986in}. For the standard parametrization of $U$ in Eq.~(\ref{eq:parametrization}), one can get the explicit expression ${\cal J} = s^{}_{12} c^{}_{12} s^{}_{23} c^{}_{23} s^{}_{13} c^2_{13} \sin \delta$. In a similar way, the other two WB invariants ${\cal I}^{}_2$ and ${\cal I}^{}_4$ can also be calculated, but the explicit analytical expressions are too lengthy to be listed here. Taking the advantage of the simple result for ${\cal I}^{}_1$ in Eq.~(\ref{eq:I1expr}), we can show that ${\cal I}^{}_1 = {\cal I}^{}_2 = {\cal I}^{}_4 = 0$ are sufficient conditions for CP conservation in the case of nondegenerate neutrino masses. First of all, ${\cal I}_1=0$ holds if and only if $\delta=0$ or $\pi$. After using ${\cal I}_1 = 0$ to eliminate the Dirac CP phase $\delta$, we can then observe that ${\cal I}_2=0$ and ${\cal I}_4=0$ give rise to two independent equations of two Majorana CP phases $\rho$ and $\sigma$, namely,
\begin{eqnarray}
&f^{}_1 \sin(2\rho) + f^{}_2 \sin(2\sigma) + f^{}_3 \sin(2\rho-2\sigma) = 0  \; ,\label{eq:I2expr} \\
&h^{}_1 \sin(2\rho) + h^{}_2 \sin(2\sigma) + h^{}_3 \sin(2\rho-2\sigma) = 0  \; ,\label{eq:I4expr}
\end{eqnarray}
where $f^{}_i$ and $h^{}_i$ (for $i = 1, 2, 3$) are functions of three mixing angles and six lepton masses. The explicit expressions of $f^{}_i$ and $h^{}_i$ can be found in Ref.~\cite{Yu:2019ihs}. Since Eqs.~(\ref{eq:I2expr}) and (\ref{eq:I4expr}) are actually nonlinear in nature, they cannot enforce $\rho$ and $\sigma$ to take only trivial values (i.e., $0$ or $\pi/2$) in general. However, it can be proved that at least in the whole physically allowed parameter space, these two equations are sufficient to ensure that $\rho$ and $\sigma$ take only trivial values ~\cite{Yu:2019ihs}, so CP conservation is justified. On the other hand, it is easy to prove that the vanishing of three invariants $\{{\cal I}_1, {\cal I}_2, {\cal I}_4 \}$ is also a necessary condition for CP conservation with nondegenerate neutrino masses~\cite{Branco:1986gr}.

In the following discussions, we shall concentrate on the partially-degenerate mass spectrum $m^{}_1 = m^{}_2 \neq m^{}_3$\footnote{The other two possibilities, i.e., $m^{}_1 \neq m^{}_2 = m^{}_3$ and $m^{}_1 = m^{}_3 \neq m^{}_2$, can be examined in a similar way.} and the completely-degenerate mass spectrum $m^{}_1 = m^{}_2 = m^{}_3$. These two special cases have not been considered in Ref.~\cite{Yu:2019ihs}.

\subsection{Partial mass degeneracy}\label{subsec:partial degeneracy}

If the partial mass degeneracy $m^{}_1 = m^{}_2 \neq m^{}_3$ is assumed, then from Eq.~(\ref{eq:I1expr}) we find that ${\cal I}_1$ vanishes automatically and it can no longer be used to investigate the properties of CP violation. Nevertheless, there exists an extra degree of freedom in the system with two degenerate neutrino masses, which can be implemented to reduce the number of CP-violating phases.

To see this point more clearly, we choose the basis where both neutrino mass matrix $M_{\nu}$ and the charged-current interaction are flavor-diagonal. In this basis, the neutrino mass matrix $M^{}_\nu = \widehat{M}^{}_\nu = {\rm diag} \{m, m, m^{}_3\}$, where we have taken $m^{}_1 = m^{}_2 =m$, is invariant under the transformation
\begin{eqnarray}
\left(\begin{matrix} \nu^{}_{1{\rm L}} \cr \nu^{}_{2{\rm L}} \cr \nu^{}_{3{\rm L}} \end{matrix}\right) \rightarrow \left(\begin{matrix} \nu^\prime_{1{\rm L}} \cr \nu^\prime_{2{\rm L}} \cr \nu^\prime_{3{\rm L}} \end{matrix}\right) = R^{\dagger}_{12}(\alpha) \left(\begin{matrix} \nu^{}_{1{\rm L}} \cr \nu^{}_{2{\rm L}} \cr \nu^{}_{3{\rm L}} \end{matrix}\right) \equiv \left(
\begin{matrix}
\cos\alpha & -\sin\alpha & 0 \cr
\sin\alpha & \cos\alpha & 0 \cr
0 & 0 & 1
\end{matrix}
\right) \left(\begin{matrix} \nu^{}_{1{\rm L}} \cr \nu^{}_{2{\rm L}} \cr \nu^{}_{3{\rm L}} \end{matrix}\right) \; ,
\label{eq:R12nu}
\end{eqnarray}
where $R^{}_{12}(\alpha)$ has been defined as the rotation matrix in the $(1, 2)$-plane with $\alpha$ being an arbitrary real rotation angle. To keep the flavor-diagonal charged-current interaction unchanged, one thus has to make the same transformation on the left-handed charged leptons simultaneously, i.e.,
\begin{eqnarray}
\left(\begin{matrix} e^{}_{\rm L} \cr \mu^{}_{\rm L} \cr \tau^{}_{\rm L}\end{matrix}\right) \rightarrow \left(\begin{matrix} e^\prime_{\rm L} \cr \mu^\prime_{\rm L} \cr \tau^\prime_{\rm L}\end{matrix}\right) = R^{\dagger}_{12}(\alpha) \left(\begin{matrix} e^{}_{{\rm L}} \cr \mu^{}_{{\rm L}} \cr \tau^{}_{{\rm L}} \end{matrix}\right) \equiv \left(
\begin{matrix}
\cos\alpha & -\sin\alpha & 0 \cr
\sin\alpha & \cos\alpha & 0 \cr
0 & 0 & 1
\end{matrix}
\right) \left(\begin{matrix} e^{}_{{\rm L}} \cr \mu^{}_{{\rm L}} \cr \tau^{}_{{\rm L}} \end{matrix}\right) \; .
\label{eq:R12l}
\end{eqnarray}
Under these transformations, the whole effective Lagrangian in Eq.~(\ref{eq:effectivelag}) is not modified except for the charged-lepton mass matrix $M^{}_l$, which together with $H^{}_l \equiv M^{}_l M^\dagger_l$ transforms as follows
\begin{eqnarray}\label{eq:Hlprime}
M^{}_l \rightarrow M_l^{\prime} = R_{12}^{\dagger}(\alpha) M^{}_{l} \;, \qquad H^{}_l \rightarrow H_l^{\prime} = R_{12}^{\dagger}(\alpha) H^{}_{l} R^{}_{12}(\alpha) \; .
\end{eqnarray}
In the chosen basis, only the charged-lepton mass matrix $M^{}_l$ is complex and thus contains all the information about CP-violating phases. Furthermore, to remove the unphysical phases related to the right-handed charged-lepton fields, we consider only the Hermitian matrix $H^{}_l$. Now we show that the rotation matrix $R^{}_{12}(\alpha)$ can be utilized to reduce the number of CP phases in the original Lagrangian. To be explicit, we directly establish the correspondence between the matrix elements of $H^{\prime}_l$ and those of $H^{}_l$, i.e.,
\begin{eqnarray}
\left\{
\begin{aligned}
& H_{11}^{\prime} = \frac{1}{2} \left[ H^{}_{11} + H^{}_{22} + (H^{}_{11} - H_{22}) \cos 2\alpha + 2 H^{}_{12} \sin 2\alpha \cos \phi^{}_{12} \right] \; , \\
& H_{22}^{\prime} = \frac{1}{2} \left[ H^{}_{11} + H^{}_{22} - (H^{}_{11} - H^{}_{22}) \cos 2\alpha - 2H^{}_{12} \sin 2\alpha \cos \phi^{}_{12} \right] \; , \\
& H_{33}^{\prime} = H^{}_{33} \; , \\
& H_{12}^{\prime} e^{{\rm i}\phi_{12}^{\prime}} = H^{}_{12} e^{{\rm i} \phi^{}_{12}} \cos^2 \alpha - \frac{1}{2} (H^{}_{11} - H^{}_{22}) \sin 2\alpha - H^{}_{12} e^{-{\rm i} \phi^{}_{12}} \sin^2 \alpha \; , \\
& H_{13}^{\prime} e^{{\rm i} \phi_{13}^{\prime}} = H^{}_{13} e^{{\rm i} \phi^{}_{13}} \cos\alpha + H^{}_{23} e^{{\rm i}\phi^{}_{23}} \sin \alpha \; , \\
& H_{23}^{\prime} e^{{\rm i} \phi_{23}^{\prime}} = H^{}_{23} e^{{\rm i} \phi^{}_{23}} \cos \alpha - H^{}_{13} e^{{\rm i} \phi^{}_{13}} \sin \alpha \; , \\
\end{aligned}
\right.
\label{eq:HprimeH}
\end{eqnarray}
where $H^{}_{ij} \equiv |(H^{}_l)^{}_{ij}|$ and $\phi^{}_{ij} \equiv \arg\left[(H^{}_l)^{}_{ij}\right]$ have been defined for $H^{}_l$ (for $i, j = 1, 2, 3$), and likewise $H^{\prime}_{ij} \equiv |(H^\prime_l)^{}_{ij}|$ and $\phi^{\prime}_{ij} \equiv \arg\left[(H^\prime_l)^{}_{ij}\right]$ for $H^\prime_l$. Note that $H^{}_l$ (or $H^\prime_l$) is Hermitian, so only three phases $\{\phi^{}_{12}, \phi^{}_{13}, \phi^{}_{23}\}$ in $H^{}_l$ (or $\{\phi^\prime_{12}, \phi^\prime_{13}, \phi^\prime_{23}\}$ in $H^\prime_l$) are independent. In the case of nondegenerate neutrino masses, where these three phases are all physical, three conditions ${\cal I}^{}_1 = 0$, ${\cal I}^{}_2 = 0$ and ${\cal I}^{}_4 = 0$ are needed to guarantee CP conservation. In the presence of mass degeneracy $m^{}_1 = m^{}_2$, we can adjust the rotation angle $\alpha$ to eliminate one phase in $H_l^{\prime}$. For example, if we set
\begin{eqnarray}
\tan\alpha = -\frac{H^{}_{13} \sin \phi^{}_{13}}{H^{}_{23} \sin\phi^{}_{23}} \; ,
\label{eq:alpha}
\end{eqnarray}
then one can immediately verify that $\phi^\prime_{13} = 0$ holds or equivalently that $(H^\prime_l)^{}_{13}$ is real, with the help of Eq.~(\ref{eq:alpha}). This is true for the most general case of $H^{}_{23} \sin \phi^{}_{23} \neq 0$. In the special case of $H^{}_{23} \sin \phi^{}_{23} = 0$, we can observe from Eq.~(\ref{eq:alpha}) that $H^\prime_{23} \sin \phi^\prime_{23} = - H^{}_{13} \sin \phi^{}_{13} \sin \alpha$ and $H^\prime_{13} \sin \phi^\prime_{13} = H^{}_{13} \sin \phi^{}_{13} \cos \alpha$, so it is possible to eliminate $\phi^\prime_{23}$ or $\phi^\prime_{13}$ by setting $\alpha = 0$ or $\pi/2$.

In general, we are left with only two phases $\{\phi^\prime_{12}, \phi^\prime_{23}\}$ in $H^\prime_l$, while $M^\prime_{\nu}$ is real and diagonal. Therefore, we can prove that only two WB invariants are needed to ensure CP conservation in the leptonic sector, which will be taken to be $\{{\cal I}^{}_2$, ${\cal I}^{}_3\}$. It is worth stressing that the choice of two independent WB invariants is by no means unique, and $\{{\cal I}^{}_2, {\cal I}^{}_3\}$ are chosen just for illustration. The proof is as follows.
\begin{itemize}
\item Now that $M^\prime_{\nu} = {\rm diag}\{m, m, m^{}_3\}$ is real and diagonal, we can directly compute the WB invariant ${\cal I}^{}_2$ in Eq.~(\ref{eq:I2}) with $\phi^\prime_{13} = 0$ in $H^\prime_l$. The analytical expression turns out to be quite simple, namely,
\begin{eqnarray}
{\cal I}^{}_2 = m m^{}_3 (m_3^2 - m^2) H^{\prime 2}_{23} \sin2\phi^\prime_{23} \; ,
\label{eq:I2m12}
\end{eqnarray}
so ${\cal I}_2=0$ leads to $\phi^\prime_{23} = 0$ or $\phi^\prime_{23} = \pi/2$. In both cases, one can find that the WB invariant ${\cal I}^{}_3$ depends on the phase $\phi^\prime_{12}$. More explicitly, for $\phi^\prime_{23} = 0$, we have ${\cal I}^{}_3 \propto \sin \phi^\prime_{12}$; while for $\phi^\prime_{23} = \pi/2$, we get ${\cal I}^{}_3 \propto \cos \phi^\prime_{12}$. As a consequence, together with ${\cal I}^{}_3 = 0$, ${\cal I}^{}_2 = 0$ implies that $\{\phi^\prime_{12} = 0,  \phi^\prime_{23} =0, \phi^\prime_{13} = 0\}$ or $\{\phi^\prime_{12} = \pi/2, \phi^\prime_{23} = \pi/2, \phi^\prime_{13}=0\}$. In either case, these trivial phases are expected for the absence of CP violation.

\item On the other hand, one can relate the CP-violating phases in $H^\prime_l$ to those in the PMNS matrix $U$. In the chosen basis, we have $M^\prime_l = U^{\dagger} \widehat{M}^{}_l$ and $H^\prime_l = U^{\dagger} \widehat{D}^{}_l U^{}$, where $\widehat{M}_l \equiv {\rm diag}\{m^{}_e, m^{}_{\mu}, m^{}_{\tau}\}$ and $\widehat{D}^{}_l \equiv \widehat{M}^2_l = {\rm diag}\{m^2_e, m^2_{\mu}, m^2_{\tau}\}$. Then it is possible to relate the three phases in $H^\prime_l$ to the three physical phases in the PMNS matrix,
\begin{eqnarray}
H^\prime_{12} e^{{\rm i}\phi^\prime_{12}} &=& \left[\left(s^2_{12} e^{{\rm i}\delta} - c^2_{12} e^{-{\rm i}\delta}\right) s^{}_{13} s^{}_{23} c^{}_{23} + s^{}_{12} c^{}_{12} (s^2_{23} - c^2_{23})\right]\Delta^{}_{\mu\tau} e^{-{\rm i}(\rho - \sigma)} \nonumber \\
&& + (\Delta^{}_{e\mu}  s^2_{23} - \Delta^{}_{\tau e} c^2_{23}) s^{}_{12} c^{}_{12} c^2_{13} e^{-{\rm i}(\rho - \sigma)} \; , \nonumber \\
H^\prime_{13} e^{{\rm i}\phi^\prime_{13}} &=& (\Delta^{}_{e\mu} s^2_{23} -\Delta^{}_{\tau e} c^2_{23}) c^{}_{12} s^{}_{13} c^{}_{13} e^{-{\rm i}(\rho+\delta)} - \Delta^{}_{\mu\tau} s^{}_{12} c^{}_{13} s^{}_{23} c^{}_{23} e^{-{\rm i}\rho} \; ,\nonumber \\
H^\prime_{23} e^{{\rm i}\phi^\prime_{23}} &=& (\Delta^{}_{e\mu} s^2_{23} -\Delta^{}_{\tau e} c^2_{23}) s^{}_{12} s^{}_{13} c^{}_{13} e^{-{\rm i}(\sigma+\delta)} + \Delta^{}_{\mu\tau} c^{}_{12} c^{}_{13} s^{}_{23} c^{}_{23} e^{-{\rm i}\sigma} \; , \nonumber
\end{eqnarray}
where it is interesting to observe that the expression of $H^\prime_{23} e^{{\rm i}\phi^\prime_{23}}$ can be obtained from that of $H^\prime_{13} e^{{\rm i}\phi^\prime_{13}}$ by simply replacing $\theta^{}_{12}$ with $\theta^{}_{12} - \pi/2$ and $\rho$ with $\sigma$. Then $\{\phi^\prime_{12} = \phi^\prime_{13} = \phi^\prime_{23} = 0\}$ or $\{\phi^\prime_{12} = \phi^\prime_{23} = \pi/2, \phi^\prime_{13} = 0\}$ is equivalent to $\{\delta = \rho = \sigma = 0\}$ or $\{\delta = \rho = 0, \sigma=\pi/2\}$, which is equivalent to CP conservation.\footnote{It should be noted that $\rho$ and $\sigma$ are Majorana-type CP-violating phases and the CP symmetry is still conserved when they take the value of $\pi/2$ in the standard parametrization in Eq.~(\ref{eq:parametrization}). The properties of three phases in $H^{}_l$ are quite different. For instance, if one of three phases in $H^{}_l$ takes the value of $\pi/2$ and the other two are zero, then the CP symmetry is violated.} This completes the proof that $\{{\cal I}^{}_2 = 0, {\cal I}^{}_3 = 0\}$ constitute the sufficient and necessary conditions of CP conservation in the case of partial mass degeneracy.
\end{itemize}

In summary, for the partial degeneracy of neutrino masses $m^{}_1 = m^{}_2 \neq m^{}_3$, there are only two independent CP-violating phases, and the vanishing of two WB invariants in Eqs.~(\ref{eq:I2}) and (\ref{eq:I3}), namely, ${\cal I}^{}_2 = 0$ and ${\cal I}^{}_3 = 0$ , serves as the sufficient and necessary conditions for the leptonic CP conservation. In addition, it is worthwhile to notice that the freedom associated with the mass degeneracy $m^{}_1 = m^{}_2$ can be implemented to reduce the number of CP-violating phases by one, leaving three flavor mixing angles intact.

\subsection{Complete mass degeneracy}\label{subsec:complete degeneracy}

If neutrino masses are completely degenerate, i.e., $m^{}_1 = m^{}_2 = m^{}_3 \equiv m$, then it is straightforward to verify that the WB invariants ${\cal I}^{}_1$, ${\cal I}^{}_2$ and ${\cal I}^{}_4$ automatically vanish, whereas ${\cal I}^{}_3$ is generally nonzero. However, compared to the case of partial mass degeneracy, the complete mass degeneracy allows for more degrees of freedom, which can be utilized to reduce the number of physical CP-violating phases.

In the same way as for the partial mass degeneracy, working in the basis where the neutrino mass matrix $M^{}_\nu = \widehat{M}^{}_\nu = {\rm diag}\{m, m, m\}$ is real and diagonal, we can introduce two successive rotations in the flavor basis
\begin{eqnarray}
\left(\begin{matrix} \nu^{}_{1{\rm L}} \cr \nu^{}_{2{\rm L}} \cr \nu^{}_{3{\rm L}} \end{matrix}\right) &\rightarrow& \left(\begin{matrix} \nu^\prime_{1{\rm L}} \cr \nu^\prime_{2{\rm L}} \cr \nu^\prime_{3{\rm L}} \end{matrix}\right) = \left[R^{}_{12}(\alpha) R^{}_{13}(\beta)\right]^{\dagger} \left(\begin{matrix} \nu^{}_{1{\rm L}} \cr \nu^{}_{2{\rm L}} \cr \nu^{}_{3{\rm L}} \end{matrix}\right) \; , ~~ \\
\left(\begin{matrix} e^{}_{{\rm L}} \cr \mu^{}_{{\rm L}} \cr \tau^{}_{{\rm L}} \end{matrix}\right) &\rightarrow& \left(\begin{matrix} e^\prime_{{\rm L}} \cr \mu^\prime_{{\rm L}} \cr \tau^\prime_{{\rm L}} \end{matrix}\right) = \left[R^{}_{12}(\alpha) R^{}_{13}(\beta)\right]^{\dagger} \left(\begin{matrix} e^{}_{{\rm L}} \cr \mu^{}_{{\rm L}} \cr \tau^{}_{{\rm L}} \end{matrix}\right) \; , \quad
\label{eq:R1213nu}
\end{eqnarray}
where the rotation matrices are defined as
\begin{eqnarray}
R^{}_{12}(\alpha) = \left(
\begin{matrix}
\cos\alpha & -\sin\alpha & 0 \cr
\sin\alpha & \cos\alpha & 0 \cr
0 & 0 & 1
\end{matrix}
\right) \; , \quad
R^{}_{13}(\beta) = \left(
\begin{matrix}
\cos\beta & 0 & \sin\beta \cr
0 & 1 & 0 \cr
-\sin\beta & 0 & \cos\beta
\end{matrix} \right) \; , \nonumber
\end{eqnarray}
with $\alpha$ and $\beta$ being two arbitrary real rotation angles. After these rotations, the neutrino mass matrix $M^{}_\nu$ is unchanged and the charged-current interaction remains to be flavor-diagonal, but the charged-lepton mass matrix $H^{}_l \equiv M^{}_l M^\dagger_l$ transforms as below
\begin{eqnarray}
H^{}_l \rightarrow H_{l}^{\prime} = \left[R^{}_{12}(\alpha) R^{}_{13}(\beta) \right]^{\dagger} \cdot H^{}_l \cdot \left[R^{}_{12}(\alpha) R^{}_{13}(\beta) \right] \; ,
\label{eq:Hlprime1213}
\end{eqnarray}
which contains all the physical CP-violating phases.

Similar to what we have done in Sec.~\ref{subsec:partial degeneracy}, we can show how to adjust $\alpha$ and $\beta$ to eliminate two CP-violating phases in $H_l^{\prime}$. This is equivalent to the reduction of the total number of CP-violating phases in the leptonic sector by two. After some straightforward calculations, we find that if $\alpha$ and $\beta$ are taken to be
\begin{eqnarray}
\tan\alpha = -\frac{H^{}_{13} \sin\phi^{}_{13}}{H^{}_{23}\sin\phi^{}_{23}} \; , \quad
\tan\beta = \frac{H^{}_{23} \sin\phi^{}_{23}}{H^{}_{12} \sin\phi^{}_{12}} \frac{1}{\cos\alpha} \; ,
\label{eq:alphabeta}
\end{eqnarray}
then $\sin\phi^\prime_{13} = \sin\phi^\prime_{23} = 0$, indicating that the imaginary parts of the matrix elements $(H_{l}^{\prime})^{}_{13}$ and $(H_{l}^{\prime})^{}_{23}$ vanish. As a result, one needs only one vanishing WB invariant, e.g., ${\cal I}^{}_3 = 0$, to eliminate the remaining one CP-violating phase in $H_l^{\prime}$. After setting $\phi^\prime_{13} = \phi^\prime_{23} = 0$ or $\pi$, we can greatly simplify the explicit expression of ${\cal I}^{}_3$, namely,
\begin{eqnarray}
{\cal I}^{}_3 = -48 {\rm i} H_{12}^{\prime 3} m^6 \left[ H^\prime_{13} H^\prime_{23} (H^\prime_{22} - H^\prime_{11}) + H^\prime_{12} (H_{13}^{\prime 2} - H_{23}^{\prime 2}) \cos\phi^\prime_{12} \right] \sin^3 \phi^\prime_{12} \; ,
\label{eq:I3alphabeta}
\end{eqnarray}
implying that ${\cal I}^{}_3 = 0$ gives rise to $\phi^\prime_{12} = 0$ or $\pi$ if the whole coefficient in front of $\sin^3\phi^\prime_{12}$ is not fine-tuned to be zero. Therefore, ${\cal I}^{}_3 = 0$ is the sufficient and necessary condition for CP conservation in the case of complete neutrino mass degeneracy, which is consistent with the conclusion previously drawn in Ref.~\cite{Branco:1998bw}.

Generally speaking, the neutrino mass matrix $M^{}_\nu = \widehat{M}^{}_\nu \equiv {\rm diag}\{m, m, m\}$ in the case of complete mass degeneracy is invariant under an arbitrary orthogonal rotation with three rotation angles. One may wonder whether it is possible to eliminate all three CP-violating phases. Now we demonstrate that this is impossible. To this end, we first carry out the most general orthogonal rotation in the flavor basis
\begin{eqnarray}
\left(\begin{matrix} \nu^{}_{1{\rm L}} \cr \nu^{}_{2{\rm L}} \cr \nu^{}_{3{\rm L}} \end{matrix}\right) &\rightarrow& \left(\begin{matrix} \nu^\prime_{1{\rm L}} \cr \nu^\prime_{2{\rm L}} \cr \nu^\prime_{3{\rm L}} \end{matrix}\right) = \left[R^{}_{12}(\alpha) R^{}_{13}(\beta) R^{}_{23}(\gamma)\right]^{\dagger} \left(\begin{matrix} \nu^{}_{1{\rm L}} \cr \nu^{}_{2{\rm L}} \cr \nu^{}_{3{\rm L}} \end{matrix}\right) \; ,
\label{eq:R121323n} \\
\left(\begin{matrix} e^{}_{{\rm L}} \cr \mu^{}_{{\rm L}} \cr \tau^{}_{{\rm L}} \end{matrix}\right) & \rightarrow & \left(\begin{matrix} e^\prime_{{\rm L}} \cr \mu^\prime_{{\rm L}} \cr \tau^\prime_{{\rm L}} \end{matrix}\right) = \left[R^{}_{12}(\alpha) R^{}_{13}(\beta) R^{}_{23}(\gamma)\right]^{\dagger} \left(\begin{matrix} e^{}_{{\rm L}} \cr \mu^{}_{{\rm L}} \cr \tau^{}_{{\rm L}} \end{matrix}\right) \; , \quad
\label{eq:R121323l}
\end{eqnarray}
where $R^{}_{12}(\alpha)$ and $R^{}_{13}(\beta)$ are the same as before and
\begin{equation}
R^{}_{23}(\gamma)=
\left(
\begin{matrix}
1 & 0 & 0 \cr
0 & \cos\gamma & -\sin\gamma \cr
0 & \sin\gamma & \cos\gamma
\end{matrix}
\right) \; , \nonumber
\end{equation}
with $\gamma$ being another real arbitrary rotation angle. Such transformations will keep the neutrino mass matrix and the charged-current interaction unchanged. However, it is straightforward to prove that the third degree of freedom can only be used to eliminate a flavor mixing angle rather than the remaining CP-violating phase. One can accomplish the proof by contradiction. First of all, given Eq.~(\ref{eq:R121323l}), the Hermitian matrix $H^{}_l$ transforms as
\begin{eqnarray}
H^{}_l \rightarrow H_{l}^{\prime} = \left[ R^{}_{12}(\alpha) R^{}_{13}(\beta) R^{}_{23}(\gamma) \right]^{\dagger} \cdot H^{}_l \cdot \left[ R^{}_{12}(\alpha) R^{}_{13}(\beta) R^{}_{23}(\gamma)\right] \; ,
\label{eq:Hl}
\end{eqnarray}
and we suppose that all three CP-violating phases in $H^\prime_l$ can be made trivial (i.e., $\phi^\prime_{12}, \phi^\prime_{23}, \phi^\prime_{13} = 0$ or $\pi$) by adjusting the rotation angles $\alpha$, $\beta$ and $\gamma$. If this is possible, then we can observe that the imaginary parts of three off-diagonal elements of $H_{l}^{\prime}$ in Eq.~(\ref{eq:Hl}) should vanish, i.e.,
\begin{eqnarray}
\left( \begin{matrix} H^\prime_{12} \sin \phi^\prime_{12} \cr H^\prime_{23} \sin \phi^\prime_{23} \cr H^\prime_{13} \sin \phi^\prime_{13} \end{matrix} \right) = \left( \begin{matrix} c^{}_\beta c^{}_\gamma & s^{}_\alpha s^{}_\gamma + c^{}_\alpha s^{}_\beta c^{}_\gamma & c^{}_\alpha s^{}_\gamma - s^{}_\alpha s^{}_\beta c^{}_\gamma \cr - c^{}_\beta s^{}_\gamma & s^{}_\alpha c^{}_\gamma - c^{}_\alpha s^{}_\beta s^{}_\gamma & c^{}_\alpha c^{}_\gamma + s^{}_\alpha s^{}_\beta s^{}_\gamma \cr -s^{}_\beta & c^{}_\alpha c^{}_\beta & - s^{}_\alpha c^{}_\beta \end{matrix} \right) \left( \begin{matrix} H^{}_{12} \sin \phi^{}_{12} \cr H^{}_{23} \sin \phi^{}_{23} \cr H^{}_{13} \sin \phi^{}_{13} \end{matrix} \right) = {\bf 0} \; ,
\label{eq:homoeq}
\end{eqnarray}
which is a system of homogeneous linear equations for $H^{}_{12} \sin \phi^{}_{12}$, $H^{}_{23} \sin \phi^{}_{23}$ and $H^{}_{13} \sin \phi^{}_{13}$. Note that $s^{}_\alpha \equiv \sin \alpha$ and $c^{}_\alpha \equiv \cos \alpha$ have been defined in Eq.~(\ref{eq:homoeq}), and likewise for $\beta$ and $\gamma$. It is interesting to notice that the determinant of the $3\times 3$ coefficient matrix in the middle of Eq.~(\ref{eq:homoeq}) is actually $-1$, which is independent of $\alpha$, $\beta$ and $\gamma$. Therefore, Eq.~(\ref{eq:homoeq}) holds if and only if $\sin\phi^{}_{12} = \sin\phi^{}_{23} = \sin\phi^{}_{13} = 0$, which runs into the contradiction with the fact that there are in general three CP-violating phases in $H^{}_l$. So this proves that even in the limit of complete mass degeneracy, there is still one nonvanishing phase so that CP can be violated in the leptonic sector, which is consistent with the conclusion drawn in Ref.~\cite{Branco:1998bw}.

In order to use the third degree of freedom to eliminate a flavor mixing angle, we can first choose the rotation angles $\alpha$ and $\beta$ to obtain $\phi_{23}^{\prime} = 0$ and $\phi_{12}^{\prime} = \phi_{13}^{\prime}$, then $H_{l}^{\prime}$ can be explicitly written as
\begin{eqnarray}
H^\prime_l = P^{}_l \left( \begin{matrix} H^\prime_{11} & H^\prime_{12} & H^\prime_{13} \cr H^\prime_{12} & H^\prime_{22} & H^\prime_{23} \cr H^\prime_{13} & H^\prime_{23} & H^\prime_{33} \end{matrix} \right) P^\dagger_l = \left(P^{}_l O^{}_l\right) \cdot \left( \begin{matrix} m^2_e & 0 & 0 \cr 0 & m^2_\mu & 0 \cr 0 & 0 & m^2_\tau \end{matrix} \right) \cdot \left(P^{}_l O^{}_l\right)^\dagger \; ,
\label{eq:PO}
\end{eqnarray}
where $P^{}_l \equiv {\rm diag}\{e^{{\rm i}\phi^\prime_{12}}, 1, 1\}$ and $O^{}_l$ is the $3\times 3$ orthogonal matrix that can be used to diagonalize the real and symmetric matrix $P^\dagger_l H^\prime_l P^{}_l$. Since the neutrino mass matrix is already diagonal, the PMNS matrix is simply given by $U = (P^{}_l O^{}_l)^\dagger = O^{\rm T}_l P^*_l$. Furthermore, noticing that the mass eigenstates $\nu^{}_2$ and $\nu^{}_3$ are now degenerate in mass and their Majorana CP phases are both vanishing, we are allowed to rotate away one mixing angle by choosing the particular parametrization of $O^{}_l$, as explicitly shown in Ref.~\cite{Mei:2003gu}.

To summarize, in the case of complete mass degeneracy, we are left with one CP-violating phase and two mixing angles. This should be compared with the case of partial mass degeneracy, where two CP-violating phases and three mixing angles are retained. It is worth mentioning that Ref.~\cite{Mei:2003gu} examines the case where both neutrino masses and the associated Majorana CP-violating phases are partially or completely degenerate at the same time, which is quite different from the scenario under consideration. It is physically inequivalent to assume the equality of two Majorana CP phases before or after the elimination of one mixing angle.

\subsection{The massless limit}

Current neutrino oscillation data still permit the lightest neutrino to be massless, so we give a brief comment on this particular situation. Without loss of generality, we take $m^{}_1 = 0$ in the case of normal neutrino mass ordering (i.e., $m^{}_1 < m^{}_2 < m^{}_3$).

If $m^{}_1 = 0$ holds, then the Majorana CP-violating phase $\rho$ associated with the mass eigenstate $\nu^{}_1$ automatically disappears from the theory. In this case, we are left with two CP-violating phases and need to require two WB invariants to vanish in order to ensure CP conservation. Luckily, none of the previously introduced four WB invariants ${\cal I}^{}_i$ (for $i=1, 2, 3, 4$) vanishes in the limit of $m^{}_1 = 0$, so we can choose any two of them to guarantee CP conservation. To be more concrete, we take the set $\{{\cal I}^{}_1, {\cal I}^{}_2 \}$. First, we can use ${\cal I}^{}_1 = 0$ to eliminate the Dirac CP-violating phase $\delta$, since ${\cal I}^{}_1$ is proportional to $\sin\delta$. Now that both $\delta$ and $\rho$ are set to be zero, ${\cal I}^{}_2$ turns out to be proportional to $\sin2\sigma$, where $\sigma$ denotes the remaining Majorana CP-violating phase. Then, the condition ${\cal I}^{}_2 = 0$ enforces $\sigma$ to take only trivial values (i.e., $0$ or $\pi/2$), implying CP conservation.

Therefore, in the limit $m^{}_1 = 0$, the vanishing of the set of two WB invariants $\{{\cal I}_1,{\cal I}_2 \}$ serves as the sufficient and necessary condition for CP conservation. As we shall see in the next section, the lightest neutrino is indeed massless at the tree level in the minimal seesaw model, which leads to the massless limit of the low-energy effective theory under consideration.

\subsection{Renormalization-group running}

In this subsection, we derive the renormalization-group equations (RGEs) of the WB invariants in leptonic sector in the effective theory, which have rarely been investigated in the literature.\footnote{The renormalization-group evolution of the WB invariants in the quark sector has been discussed in Ref.~\cite{Feldmann:2015nia}.} These RGEs can be applied to examine the evolution of the WB invariants and establish the connection between CP violation at low- and high-energy scales. At the one-loop level, the evolution of the effective Majorana neutrino mass matrix $M^{}_\nu$ and the charged-lepton mass matrix $M^{}_l$ are governed by the following RGEs~\cite{Chankowski:1993tx, Babu:1993qv, Antusch:2001ck, Antusch:2002rr, Antusch:2003kp, Mei:2003gn, Antusch:2005gp, Mei:2005qp, Luo:2005sq, Xing:2005fw, Xing:2011zza, Ohlsson:2012pg, Ohlsson:2013xva}
\begin{eqnarray}
\frac{{\rm d} M^{}_{\nu}}{{\rm d}t} & = & \alpha^{}_{\nu} M^{}_{\nu} -\frac{3}{2} \left[ \left(Y^{}_l Y_l^{\dagger} \right) M^{}_{\nu} + M^{}_{\nu} \left( Y^{}_l Y_l^{\dagger} \right)^{\rm T} \right] \; , \label{eq:RGEMnu}\\
\frac{{\rm d}M^{}_l}{{\rm d}t} & = & \alpha^{}_l M^{}_l + \frac{3}{2} \left( Y^{}_l Y_l^{\dagger} \right) M^{}_l \; , \label{eq:RGEMl}
\end{eqnarray}
where $t\equiv {\rm ln}(\mu/\Lambda^{}_{\rm EW})/(16 \pi^2)$ has been defined with $\Lambda^{}_{\rm EW}$ being the electroweak scale and $\mu$ being the renormalization scale between $\Lambda^{}_{\rm EW}$ and the seesaw scale. In the SM framework, we have $\alpha^{}_{\nu} \approx -3g^2_2 + \lambda + 6 y^2_t$ and $\alpha^{}_l \approx -9g^2_1/4 - 9g^2_2/4 + 3y^2_t$, where $g^{}_1$ and $g^{}_2$ are the SM gauge couplings, $y^{}_t$ the top-quark Yukawa coupling, and $\lambda$ the quartic Higgs coupling~\cite{Xing:2011zza}.

Starting with Eqs.~(\ref{eq:RGEMnu})-(\ref{eq:RGEMl}) and recalling the definitions of $H^{}_l \equiv M^{}_l M^\dagger_l$, $H^{}_\nu \equiv M^{}_\nu M^\dagger_\nu$ and $G^{}_{l\nu} \equiv M^{}_\nu H^*_l M^\dagger_\nu$, one can easily find
\begin{eqnarray}
\frac{{\rm d} H^{}_l}{{\rm d}t} & = & 2\alpha^{}_l H^{}_l + 6 H_l^2/v^2 \; ,
\label{eq:RGEHl} \\
\frac{{\rm d} H^{}_{\nu}}{{\rm d}t} & = & 2 \alpha_{\nu} H^{}_{\nu} - 3 (H ^{}_l H^{}_{\nu} + H ^{}_\nu H^{}_l)/v^2 - 6 G^{}_{l\nu}/v^2 \; , \label{eq:RGEHnu} \\
\frac{{\rm d}G^{}_{l\nu}}{{\rm d}t} & = & 2(\alpha^{}_{\nu} + \alpha^{}_{l}) G^{}_{l\nu} -3(G^{}_{l\nu} H^{}_l + H^{}_l G^{}_{l\nu})/v^2 \; , \label{eq:RGEGlnu}
\end{eqnarray}
where the relation $Y^{}_l = \sqrt{2} M^{}_l/v$ has been used. It is then straightforward to calculate the RGEs of the WB invariants ${\cal I}^{}_i$ (for $i = 1, 2, 3, 4$). The final results are summarized as follows.
\begin{itemize}
\item First, as shown in Eq.~(\ref{eq:I1expr}), ${\cal I}^{}_1$ is proportional to the Jarlskog invariant ${\cal J}$, which depends only on the Dirac CP-violating phase $\delta$ in the standard parametrization of the PMNS matrix. For this WB invariant, we have
\begin{eqnarray}\label{eq:RGEI1}
\frac{{\rm d}{\cal I}^{}_1}{{\rm d}t} ={\rm Tr} \left\{ \frac{{\rm d}}{{\rm d}t}\left[H^{}_\nu, H^{}_l \right]^3\right\} = 6(\alpha_{\nu}^{} + \alpha_{l}^{}) {\cal I}^{}_{1} + 9 {\cal I}^{(1)}_{1}/v^2 - 18 {\cal I}^{(2)}_{1}/v^2 \; ,
\end{eqnarray}
where ${\cal I}^{(1)}_{1} \equiv {\rm Tr} \left\{ \left[H^{}_\nu, H^{}_l \right]^2 \cdot\left[H^{}_\nu, H^{2}_l \right] \right\}$ and ${\cal I}^{(2)}_{1} \equiv {\rm Tr} \left\{ \left[H^{}_\nu, H^{}_l \right]^2 \cdot\left[G^{}_{l \nu}, H^{}_l \right] \right\}$ are also two WB invariants. Interestingly, it is easy to derive the explicit expression of ${\cal I}^{(1)}_{1}$, i.e.,
\begin{eqnarray}\label{eq:RGEI11}
{\cal I}^{(1)}_{1} = - 4 {\rm i}(m_e^2 + m_{\mu}^2 + m_{\tau}^2) \Delta^{}_{21} \Delta^{}_{31} \Delta^{}_{32} \Delta^{}_{e\mu} \Delta^{}_{\mu\tau} \Delta^{}_{\tau e} {\cal J} = \frac{2}{3} (m_e^2 + m_{\mu}^2 + m_{\tau}^2) {\cal I}_1 \; ,
\end{eqnarray}
which is proportional to ${\cal I}^{}_1$ itself and thus to the Jarlskog invariant ${\cal J}$. In the derivation of Eq.~(\ref{eq:RGEI11}), we have made use of the identities ${\rm Tr} \left\{ \left[H^{}_\nu, H^{}_l \right]^2 \cdot\left[H^{}_\nu, H^{2}_l \right] \right\} = 2{\rm Tr} \left\{ \left[H^{}_\nu, H^{}_l \right]^3 H^{}_l\right\}$ and ${\rm Tr} \left\{ \left[H^{}_\nu, H^{}_l \right]^3 H^{}_l\right\} = {\rm Tr} \left\{ \left[H^{}_\nu, H^{}_l \right]^3 \right\} \cdot {\rm Tr}\left( H^{}_l\right)/3$. However, the WB invariant ${\cal I}^{(2)}_{1}$ depends on all three CP-violating phases in the PMNS matrix, i.e., $\{\delta, \rho, \sigma\}$, and its explicit expression turns out to be quite complicated and will be omitted here.

For illustration, let us consider the possibility to radiatively generate a nontrivial value of $\delta$ via the RGE from a vanishing $\delta$ at some high-energy scale~\cite{Luo:2005sq, Xing:2005fw, Xing:2011zza, Ohlsson:2012pg}. In this case, we set $\delta = 0$ as the initial condition, then the expression of ${\cal I}^{(2)}_{1}$ can be greatly simplified to
\begin{eqnarray}
{\cal I}^{(2)}_{1} = 2{\rm i} \{ & +& H^{}_{13} [H^{}_{13} (H^2_{12} - H^2_{23}) + H^{}_{12} H^{}_{23} (H^{}_{33} - H^{}_{11})] m^{}_1 m^{}_3 \Delta^{}_{12} \Delta^{}_{23} \sin(2\rho) \nonumber \\
&+& H^{}_{23} [ H^{}_{23} (H^2_{13} - H^2_{12}) + H^{}_{12} H^{}_{13} (H^{}_{22} - H^{}_{33})] m^{}_3 m^{}_2 \Delta^{}_{13} \Delta^{}_{12} \sin(2\sigma) \label{eq:RGEI12} \\
& +& H^{}_{12} [ H^{}_{12} (H^2_{23} - H^2_{13}) + H^{}_{23} H^{}_{13} (H^{}_{11} - H^{}_{22})] m^{}_2 m^{}_1 \Delta^{}_{23} \Delta^{}_{13} \sin(2\rho - 2\sigma)
\} \; . \qquad \nonumber
\end{eqnarray}
As $\delta$ has been set to zero, the moduli of the elements of $H^{}_l$ can be directly related to three charged-lepton masses and three flavor mixing angles via
\begin{eqnarray}
H^{}_{12} &=& s^{}_{12}c^{}_{12}c^2_{13} \Delta^{}_{e\mu} - \left[s^{}_{12}c^{}_{12} (s^2_{13} c^2_{23} - s^2_{23}) + (c^2_{12} - s^2_{12})s^{}_{13} s^{}_{23} c^{}_{23} \right] \Delta^{}_{\mu\tau} \;, \nonumber \\
H^{}_{13} &=& c^{}_{12} s^{}_{13} c^{}_{13} \Delta^{}_{e\mu} - (s^{}_{12} s^{}_{23} - c^{}_{12} s^{}_{13} c^{}_{23})c^{}_{13} c^{}_{23} \Delta^{}_{\mu\tau} \; , \nonumber \\
H^{}_{23} &=& s^{}_{12} s^{}_{13} c^{}_{13} \Delta^{}_{e\mu} + (c^{}_{12} s^{}_{23} + s^{}_{12} s^{}_{13} c^{}_{23})c^{}_{13} c^{}_{23} \Delta^{}_{\mu\tau} \; , \nonumber \\
H^{}_{11} &=& m^2_e - (1 - c^2_{12} c^2_{13})\Delta^{}_{e\mu} - (s^{}_{12} s^{}_{23} - c^{}_{12} s^{}_{13} c^{}_{23})^2 \Delta^{}_{\mu\tau} \; , \nonumber \\
H^{}_{22} &=& m^2_\mu + s^2_{12}c^2_{13}\Delta^{}_{e\mu} - (c_{12}^{} s_{23}^{} + s_{12}^{} s_{13}^{} c_{23}^{})^2 \Delta^{}_{\mu\tau} \; , \nonumber \\
H^{}_{33} &=& m^2_\tau + s^2_{13} \Delta^{}_{e\mu} + (s^2_{13} + c^2_{13} s^2_{23})\Delta^{}_{\mu\tau} \; , \nonumber
\end{eqnarray}
which can be inserted back into Eq.~(\ref{eq:RGEI12}) to obtain the explicit expression of ${\cal I}^{(2)}_1$.
From Eqs.~(\ref{eq:RGEI1})-(\ref{eq:RGEI12}), we can observe that
\begin{itemize}
\item If CP is conserved (namely, $\delta = 0$ and $\rho = \sigma = 0$ or $\pi/2$) at the initial high-energy scale, then ${\cal I}^{(1)}_1 = {\cal I}^{(2)}_1 = {\cal I}^{}_1 = 0$ and ${\rm d}{\cal I}^{}_1/{\rm d}t$ vanishes, implying that ${\cal I}^{}_1$ will stay at zero all the way down to low-energy scales.
\item If CP is violated with $\delta = 0$ but nontrivial values of $\rho$ or $\sigma$ at some high-energy scale, then ${\rm d}{\cal I}^{}_1/{\rm d}t$ is no longer vanishing, as a consequence of the nonzero ${\cal I}^{(2)}_1$ in Eq.~(\ref{eq:RGEI12}). Consequently, as the energy scale evolves, a nonzero value of ${\cal I}^{}_1$ will be developed, leading to a nonzero $\delta$. As already stressed in Ref.~\cite{Luo:2005sq}, a nontrivial value of the Dirac CP phase $\delta$ can be generated from the Majorana CP phase $\rho$ or $\sigma$ via the RG running, even though $\delta = 0$ is assumed at the beginning.
\end{itemize}

\item Then, we can derive the RGE of ${\cal I}^{}_2$ defined in Eq.~(\ref{eq:I2}) in a similar way, namely,
\begin{eqnarray}
\frac{{\rm d}{\cal I}^{}_2}{{\rm d}t} = 4(\alpha_\nu^{} + \alpha_l^{}) {\cal I}^{}_{2} - 6\ {\rm Im}\left\{ {\rm Tr} \left[ H^{}_l H^{}_\nu H^{}_l G^{}_{l\nu} + H^{}_l G^2_{l\nu}\right] \right\}/v^2 \; ,
\label{eq:RGEI2}
\end{eqnarray}
where one can easily verify that the second term on the right-hand side actually vanishes due to the hermiticity of $H^{}_l$, $H^{}_\nu$ and $G^{}_{l\nu}$ and the cyclic invariance of the trace. As an immediate consequence, the derivative of the WB invariant ${\cal I}^{}_2$ is proportional to itself. We can formally integrate Eq.~(\ref{eq:RGEI2}) and obtain
\begin{eqnarray}
{\cal I}^{}_2(t) = {\cal I}^{}_2(0) \exp\left\{4\int^t_0 \left[ \alpha^{}_\nu(t^\prime) + \alpha^{}_l(t^\prime)\right] {\rm d}t^\prime \right\} \; ,
\label{eq:I2sol}
\end{eqnarray}
where ${\cal I}^{}_2(0) \equiv {\cal I}^{}_2(t = 0)$ stands for the value at the electroweak scale $\mu = \Lambda^{}_{\rm EW}$ while ${\cal I}^{}_2(t)$ for the value at an arbitrary high-energy scale $\mu = \Lambda$. For the direct connection between low- and high-energy mass or mixing parameters in an integral form, one may be referred to previous works~\cite{Xing:2017mkx, Xing:2018kto, Zhu:2018dvj, Zhang:2020lsd}.

Since ${\cal I}^{}_2$ depends on all the three CP phases, its explicit expression is rather lengthy. As before, by setting $\delta = 0$ at some energy scale, we arrive at
\begin{eqnarray}
{\cal I}^{}_2 = H^2_{13} m^{}_1 m^{}_3 \Delta^{}_{13} \sin(2\rho) + H^2_{23} m^{}_2 m^{}_3 \Delta^{}_{23} \sin(2\sigma) + H^2_{12} m^{}_1 m^{}_2 \Delta^{}_{12} \sin(2\rho - 2\sigma) \; .
\label{eq:I2del0}
\end{eqnarray}
For ${\cal I}^{}_3$, the RGE can be calculated easily and it is interesting to find
\begin{eqnarray}
\frac{{\rm d}{\cal I}^{}_3}{{\rm d}t} =  6(\alpha^{}_\nu + 2 \alpha^{}_l) {\cal I}^{}_3 + 9 {\rm Tr}\left\{ \left[G^{}_{l\nu}, H^{}_l\right]^2 \cdot \left[G^{}_{l\nu}, H^2_l\right] \right\}/v^2 \; ,
\label{eq:RGEI3}
\end{eqnarray}
where the second term on the right-hand side is similar to ${\cal I}^{(1)}_1$ in Eq.~(\ref{eq:RGEI1}) and the difference is just to replace $H^{}_\nu$ in the latter by $G^{}_{l\nu}$. After a straightforward calculation, we can obtain
\begin{eqnarray}
{\rm Tr}\left\{ \left[G^{}_{l\nu}, H^{}_l\right]^2 \cdot \left[G^{}_{l\nu}, H^2_l\right] \right\} = 2 {\rm Tr}\left\{ \left[G^{}_{l\nu}, H^{}_l\right]^3 H^{}_l \right\} = \frac{2}{3} \left(m^2_e + m^2_\mu + m^2_\tau\right) {\cal I}^{}_3 \; ,
\end{eqnarray}
such that the RGE of ${\cal I}^{}_3$ can be formally solved as in the case of ${\cal I}^{}_2$, i.e.,
\begin{eqnarray}
{\cal I}^{}_3(t) = {\cal I}^{}_3(0) \exp\left\{3 \int^t_0 \left[2\alpha^{}_\nu(t^\prime) + 4\alpha^{}_l(t^\prime) + \sum_\alpha y^2_\alpha(t^\prime)\right] {\rm d}t^\prime \right\} \; ,
\label{eq:I3sol}
\end{eqnarray}
where $y^{}_\alpha \equiv \sqrt{2} m^{}_\alpha/v$ denotes the charged-lepton Yukawa coupling for $\alpha = e, \mu, \tau$. Since the RGEs of $\alpha^{}_\nu(t)$, $\alpha^{}_l(t)$ and $y^{}_\alpha(t)$ can be separately solved, we establish another direct connection between the high- and low-energy WB invariants.

\item Finally, let us investigate the RGE of ${\cal I}^{}_4$, which has been defined in Eq.~(\ref{eq:I4}). The final result is
\begin{eqnarray}
\frac{{\rm d}{\cal I}^{}_4}{{\rm d}t} = 2(3\alpha^{}_\nu + 2\alpha^{}_l) {\cal I}^{}_4 - 12 {\cal I}^{(1)}_4 /v^2 \; ,
\label{eq:RGEI4}
\end{eqnarray}
where ${\cal I}^{(1)}_4 \equiv {\rm Im}\left\{{\rm Tr}\left[H^{}_l H^{}_\nu G^2_{l\nu}\right]\right\}$ has been introduced. Notice that a few useful identities, i.e., ${\rm Im}\left\{{\rm Tr}\left[H^{}_l H^2_\nu H^{}_l G^{}_{l\nu}\right]\right\} = {\rm Im}\left\{{\rm Tr}\left[H^{}_l G^{}_{l\nu} H^{}_\nu G^{}_{l\nu} \right]\right\} = 0$ and ${\rm Tr}\left[H^{}_l H^{}_\nu H^{}_l H^{}_\nu G^{}_{l\nu}\right] = {\rm Tr}\left[H^{}_l H^{}_\nu G^2_{l\nu}\right]$, have been used. Because of the second term on the right-hand side of Eq.~(\ref{eq:RGEI4}), it is not possible to directly solve the RGE of ${\cal I}^{}_4$. To render the analytical formulas of ${\cal I}^{}_4$ and ${\cal I}^{(1)}_4$ readable, we set $\delta = 0$ and then get
\begin{eqnarray}
{\cal I}^{}_{4} = &+& H^2_{13} m^{}_1 m^{}_3 \Delta^{}_{13} (m_1^2 + m_3^2) \sin(2\rho) \nonumber \\ &+& H^2_{23} m^{}_2 m^{}_3 \Delta^{}_{23} (m_2^2 + m_3^2) \sin(2\sigma) \nonumber \\
&+& H^2_{12} m^{}_1 m^{}_2 \Delta^{}_{12} (m_1^2 + m_2^2) \sin(2\rho - 2\sigma) \; ,
\label{eq:I4del0}
\end{eqnarray}
and
\begin{eqnarray}
{\cal I}^{(1)}_{4} =
&+& H^{}_{13} \left[H^{}_{12} H^{}_{23} + H^{}_{13}(H^{}_{11} + H^{}_{33})\right] m^{}_1 m^{}_3 \Delta_{13} (m_1^2 + m_3^2) \sin(2\rho) \nonumber \\
&+&H^{}_{23} \left[H^{}_{12} H^{}_{13} + H^{}_{23}(H^{}_{22} + H^{}_{33}) \right] m^{}_2 m^{}_3 \Delta^{}_{23} (m_2^2 + m_3^2) \sin(2\sigma) \nonumber\\
&+& H^{}_{12} \left[H^{}_{13} H^{}_{23} + H^{}_{12} (H^{}_{11} + H^{}_{22})\right] m^{}_1 m^{}_2 \Delta^{}_{12} (m_1^2 + m_2^2) \sin(2\rho-2\sigma) \; .
\label{eq:I41del0}
\end{eqnarray}
Given ${\cal I}^{}_1 = 0$ or equivalently $\delta = 0$, we can see that ${\cal I}^{}_2$ in Eq.~(\ref{eq:I2del0}), ${\cal I}^{}_4$ and ${\cal I}^{(1)}_4$ are vanishing if $\rho$ and $\sigma$ take trivial values of $0$ or $\pi/2$ at the beginning. This is also true for ${\cal I}^{}_3$, although its expression has not been explicitly written down.
\end{itemize}

To conclude, we find that ${\rm d}{\cal I}^{}_2/{\rm d}t = 4(\alpha^{}_\nu + \alpha^{}_l) {\cal I}^{}_2$ and ${\rm d}{\cal I}^{}_3/{\rm d}t = 3 \left[ 2\alpha^{}_\nu + 4\alpha^{}_l + (y^2_e + y^2_\mu + y^2_\tau) \right] {\cal I}^{}_3$, which can be formally solved, and thus establish a direct link between low- and high-energy WB invariants. For ${\cal I}^{}_1$ and ${\cal I}^{}_4$, their derivatives with respect to $t = \left[ \ln(\mu/\Lambda^{}_{\rm EW}) \right]/(16\pi^2)$ turn out to be not proportional to themselves. However, if CP conservation is assumed at some energy scale, i.e., all the three CP phases take trivial values, then CP will be conserved all the way down to the electroweak scale. If one of three CP phases is nontrivial at the beginning, namely, CP violation exists in the theory, the other phases will be generated radiatively during the RGE running. In the case of partial or complete neutrino mass degeneracy, one can choose suitable WB invariants from $\{{\cal I}^{}_1, {\cal I}^{}_2, {\cal I}^{}_3, {\cal I}^{}_4\}$ and apply the corresponding RGEs to study their running behaviors.

\section{Canonical Seesaw Model}\label{sec:high}

The partial or complete mass degeneracy of three light neutrinos has already been excluded by neutrino oscillation data~\cite{Esteban:2020cvm, Capozzi:2018ubv}, which require two independent neutrino mass-squared differences to be $\Delta^{}_{21} \approx 7.4\times 10^{-5}~{\rm eV}^2$ and $\Delta^{}_{31} \approx \pm 2.5\times 10^{-3}~{\rm eV}^2$. On the other hand, as we have mentioned before, the effective theory considered in the previous section is valid when the heavy degrees of freedom associated with neutrino mass generation are integrated out. Therefore, we now examine the necessary and sufficient conditions for CP conservation with a partial or complete mass degeneracy of three heavy Majorana neutrinos in the canonical seesaw model, for which the gauge-invariant Lagrangian has been given in Eq.~(\ref{eq:seesawlag}). After the spontaneous gauge symmetry breaking, it can be rewritten as
\begin{eqnarray}
\mathcal{L}^{}_{\text{lepton}} = -\overline{l^{}_{\rm L}} M^{}_l l^{}_{\rm R} - \overline{\nu^{}_{\rm L}} M^{}_{\rm D} N^{}_{\rm R}  - \frac{1}{2} \overline{N^{\rm C}_{\rm R}} M^{}_{\rm R} N^{}_{\rm R} + \frac{g}{\sqrt{2}} \overline{l^{}_{\rm L}} \gamma^{\mu} \nu^{}_{\rm L}  W^{-}_{\mu} + \text{h.c.} \; ,
\label{eq:seesawlagbro}
\end{eqnarray}
where the charged-current interaction has been included to cover all the possible places for CP violation. In the presence of right-handed neutrinos, the sufficient and necessary conditions for CP conservation in the full seesaw model are equivalent to the existence of three unitary matrices $U^{}_{\rm L}$, $U^{}_{\rm R}$ and $V^{}_{\rm R}$ such that the Lagrangian in Eq.~(\ref{eq:seesawlagbro}) is invariant under
\begin{eqnarray}\label{eq:transformationSS}
l^{}_{\rm L} \to  U^{}_{\rm L} {\cal C}l^*_{\rm L} \; , \quad \nu^{}_{\rm L} \to U^{}_{\rm L} {\cal C}\nu^*_{\rm L} \; , \quad
l^{}_{\rm R} \to  U^{}_{\rm R} {\cal C}l^*_{\rm R} \; , \quad N^{}_{\rm R} \to  V^{}_{\rm R} {\cal C}N^*_{\rm R} \; , \quad
W^-_\mu \to  -(-1)^{\delta^{}_{0\mu}} W^+_\mu \; , \quad
\end{eqnarray}
where the notations are the same as in Eq.~(\ref{eq:transformation}). In terms of the fermion mass matrices, one can easily prove that this is equivalent to the conditions
\begin{eqnarray}\label{eq:necsuffss}
U^\dagger_{\rm L} M^{}_l U^{}_{\rm R} = M^*_l \; , \quad U^\dagger_{\rm L} M^{}_{\rm D} V^{}_{\rm R} =  M^*_{\rm D} \; , \quad V^{\rm T}_{\rm R} M^{}_{\rm R} V^{}_{\rm R} = - M^*_{\rm R} \; , \quad
\end{eqnarray}
which will be used to construct the WB invariants for CP conservation, similar to the construction in the effective theory. To this end, we further introduce $H^{}_{\rm D} \equiv M^\dagger_{\rm D} M^{}_{\rm D}$, $H^{}_{\rm R} \equiv M^\dagger_{\rm R} M^{}_{\rm R}$, $G^{}_{\rm DR} \equiv M^\dagger_{\rm R} H^*_{\rm D} M^{}_{\rm R}$ and
\begin{eqnarray}
H^{}_n \equiv M^\dagger_{\rm D} \left(H^{}_l\right)^n M^{}_{\rm D} \; , \quad
G^{}_n \equiv M^\dagger_{\rm R} H^*_n M^{}_{\rm R} \; ,
\label{eq:HnGn}
\end{eqnarray}
where $n$ denotes the positive integer. It is straightforward to verify that the transformation rules for these newly-defined Hermitian matrices are as follows
\begin{eqnarray}
V^\dagger_{\rm R} H^{}_{\rm D} V^{}_{\rm R} = H^*_{\rm D} \; , ~~ V^\dagger_{\rm R} H^{}_{\rm R} V^{}_{\rm R} = H^*_{\rm R} \; , ~~ V^\dagger_{\rm R} G^{}_{\rm DR} V^{}_{\rm R} = G^*_{\rm DR} \; , ~~ V^\dagger_{\rm R} H^{}_n V^{}_{\rm R} = H^*_n \; , ~~
V^\dagger_{\rm R} G^{}_n V^{}_{\rm R} = G^*_n \; , ~~
\label{eq:transGH}
\end{eqnarray}
which are universal and make the construction of WB invariants much easier. As shown in Ref.~\cite{Branco:2001pq}, the sufficient and necessary conditions for CP conservation are equivalent to the vanishing of a minimal set of WB invariants.

Before constructing the WB invariants, we count the number of physical parameters in the canonical seesaw model and will pay a particular attention to the CP phases. Without loss of generality, one can always choose the basis, in which $M^{}_l$, $M^{}_{\rm R}$ and the charged-current interaction in Eq.~(\ref{eq:seesawlagbro}) are simultaneously diagonal, so that the complex mass matrix $M^{}_{\rm D}$ will be the only source of CP violation. Following Refs.~\cite{Branco:2001pq, Endoh:2000hc}, we adopt the convenient parametrization $M^{}_{\rm D} = U^{}_{\rm D} Y^{}_\Delta$, where $U^{}_{\rm D}$ is a $3\times 3$ unitary matrix and $Y^{}_{\Delta}$ is a lower triangular matrix, i.e.,
\begin{equation}
Y^{}_{\Delta}=
\left(
\begin{array}{ccc}
y^{}_{11} & 0 & 0 \\
y^{}_{21} e^{{\rm i} \phi^{}_{21}} & y^{}_{22} & 0 \\
y^{}_{31} e^{{\rm i} \phi^{}_{31}} & y^{}_{32} e^{{\rm i} \phi^{}_{32}} & y^{}_{33} \\
\end{array}
\right) \; ,
\label{eq:YDelta}
\end{equation}
where $y^{}_{ij}$ (for $1 \leq j \leq i \leq 3$) are all real and positive parameters and $\phi^{}_{ij}$ (for $ij = 21, 31, 32$) are the phases of three off-diagonal nonzero elements. As usual, three unphysical phases of $U^{}_{\rm D}$ can be eliminated by redefining the phases of $\nu^{}_{\rm L}$, $l^{}_{\rm L}$ and $l^{}_{\rm R}$, leaving the Lagrangian unchanged. Therefore, $M^{}_{\rm D}$ contains only fifteen real parameters, six of which are phases. To be more explicit, we rewrite it as $M^{}_{\rm D} = U^{}_{\xi} P^{}_{\alpha} Y^{}_{\zeta} P^{}_{\beta}$~\cite{Branco:2001pq},
where $P^{}_{\alpha} = \text{diag} \left\{1, e^{{\rm i}\alpha^{}_1}, e^{{\rm i} \alpha^{}_2} \right\}$ and $P^{}_{\beta} = \text{diag} \left\{1, e^{{\rm i} \beta^{}_1}, e^{{\rm i}\beta^{}_2} \right\}$ are two diagonal phase matrices. In addition,
\begin{equation}\label{eq:Yzeta}
Y^{}_{\zeta} =
\left(
\begin{array}{ccc}
y^{}_{11} & 0 & 0\\
y^{}_{21} & y^{}_{22} & 0\\
y^{}_{31} & y^{}_{32} e^{{\rm i} \zeta} & y^{}_{33}\\
\end{array}
\right) \; ,
\end{equation}
is related to $Y^{}_\Delta$ by properly factorizing out relevant phases, and $U^{}_{\xi}$ is the Cabibbo-Kobayashi-Maskawa (CKM)-like unitary matrix with $\xi$ being the CP phase and three rotation angles are $\{\theta^{\rm D}_{12}, \theta^{\rm D}_{13}, \theta^{\rm D}_{23}\}$. In this way, fifteen real parameters of $M^{}_{\rm D}$ are now specified, i.e., six phases $\{\xi, \zeta, \alpha^{}_1, \alpha^{}_2, \beta^{}_1, \beta^{}_2\}$ and nine real parameters $\{\theta^{\rm D}_{12}, \theta^{\rm D}_{13}, \theta^{\rm D}_{23}\}$ and $\{y^{}_{11}, y^{}_{22}, y^{}_{33}, y^{}_{21}, y^{}_{31}, y^{}_{32} \}$. All the information about CP violation is represented by six phases of $M^{}_{\rm D}$. As we shall show soon, CP is conserved if and only if $\sin \alpha^{}_1 = \sin \alpha^{}_2 = \sin \xi = \sin\zeta = \sin 2\beta^{}_1 = \sin 2\beta^{}_2 = 0$ holds.\footnote{It is worth noticing that $\beta^{}_1$ and $\beta^{}_2$ are actually the Majorana-type CP phases, and can take the value of $\pi/2$ without violating the CP symmetry.} Using the adopted parametrization of $M^{}_{\rm D}$, we can obtain
\begin{equation}
\label{eq:HD parameterization}
H^{}_{\rm D} = M^\dagger_{\rm D} M^{}_{\rm D} = P_{\beta}^{\dagger} Y_{\zeta}^{\dagger} Y^{}_{\zeta} P^{}_{\beta} \; ,
\end{equation}
where only three phases $\{\zeta, \beta^{}_1, \beta^{}_2\}$ are involved. Consequently, even if $H^{}_{\rm D}$ was real, there would be still CP violation. Different from the effective theory, in which real $H^{}_l = M^{}_l M^\dagger_l$ implies CP conservation, all the six phases of $M^{}_{\rm D}$ are important~\cite{Branco:2001pq}.

\subsection{Nondegenerate masses}
\label{subsec:complete theory nondegerate mass}
First of all, we summarize the main results in the case of nondegenerate masses, namely, $M^{}_1\neq M^{}_2 \neq M^{}_3$, where $M^{}_i$ stands for the heavy Majorana neutrino mass (for $i = 1, 2, 3$). As already demonstrated in Ref.~\cite{Branco:2001pq}, the following six conditions
\begin{eqnarray}
\widetilde{\cal I}^{}_1 &\equiv& {\rm Im}\left\{{\rm Tr}\left[H^{}_{\rm D} H^{}_{\rm R} G^{}_{\rm DR}\right]\right\} = 0 \; ,\label{eq:J1}\\
\widetilde{\cal I}^{}_2 &\equiv& {\rm Im}\left\{{\rm Tr}\left[H^{}_{\rm D} H^{2}_{\rm R} G^{}_{\rm DR}\right]\right\} = 0 \; ,\label{eq:J2}\\
\widetilde{\cal I}^{}_3 &\equiv& {\rm Im}\left\{{\rm Tr}\left[H^{}_{\rm D} H^{2}_{\rm R} G^{}_{\rm DR} H^{}_{\rm R} \right]\right\} = 0 \; ,\label{eq:J3}\\
\widetilde{\cal I}^{}_4 &\equiv& {\rm Im}\left\{{\rm Tr}\left[H^{}_{1} H^{}_{\rm R} G^{}_{1}\right]\right\} = 0 \; , \label{eq:J4}\\
\widetilde{\cal I}^{}_5 &\equiv& {\rm Im}\left\{{\rm Tr}\left[H^{}_{1} H^{2}_{\rm R} G^{}_{1}\right]\right\} = 0 \; , \label{eq:J5}\\
\widetilde{\cal I}^{}_6 &\equiv& {\rm Im}\left\{{\rm Tr}\left[H^{}_{1} H^{2}_{\rm R} G^{}_{1} H^{}_{\rm R}\right]\right\} = 0 \; , \label{eq:J6}
\end{eqnarray}
must be fulfilled to guarantee CP conservation. Notice that $H^{}_n$ and $G^{}_n$ with $n = 1$ introduced in Eq.~(\ref{eq:HnGn}) have been used in Eqs.~(\ref{eq:J4})-(\ref{eq:J6}). With the help of the transformation rules in Eq.~(\ref{eq:transGH}), we can easily prove that $\widetilde{\cal I}^{}_i $ ($i = 1, 2, \cdots, 6$) are WB invariants and $\widetilde{\cal I}^{}_i = 0$ in Eqs.~(\ref{eq:J1})-(\ref{eq:J6}) serve as the sufficient and necessary conditions for CP conservation.

Since $\{\widetilde{\cal I}^{}_1, \widetilde{\cal I}^{}_2, \widetilde{\cal I}^{}_3\}$ depend only on three phases in $H^{}_{\rm D}$, i.e., \{$\zeta, \beta^{}_1, \beta^{}_2$\}, the vanishing of these three WB invariants gives three independent constraints on the relevant three phases. The other WB invariants $\{\widetilde{\cal I}^{}_4, \widetilde{\cal I}^{}_5, \widetilde{\cal I}^{}_6\}$ depend on all the six phases in $M^{}_{\rm D}$. After three phases in $H^{}_{\rm D}$ are eliminated by Eqs.~(\ref{eq:J1})-(\ref{eq:J3}), we are left with another set of three independent constraints from Eqs.~(\ref{eq:J4})-(\ref{eq:J6}) on the remaining three phases, i.e., \{$\xi, \alpha^{}_1, \alpha^{}_2$\}. However, as has been explained in Ref.~\cite{Yu:2019ihs}, although $\{\widetilde{\cal I}^{}_1, \widetilde{\cal I}^{}_2, \widetilde{\cal I}^{}_3\}$ are independent, the vanishing of them leads to three nonlinear equations of $\zeta$, $\beta^{}_1$ and $\beta^{}_2$, from which nontrivial solutions (i.e., other than 0 and $\pi/2$) of these three phases can be obtained for some special values of other physical parameters. For this reason, we recommend another set of three invariants $\{\widetilde{\cal I}^{}_1, \widetilde{\cal I}^{}_2, \widetilde{\cal I}_3^{\prime}\}$, where the new WB invariant $\widetilde{\cal I}_3^{\prime} \equiv {\rm Tr} \left\{ \left[ H^{}_{\rm R}, H^{}_{\rm D} \right]^3 \right\}$ replaces the original one $\widetilde{\cal I}^{}_3$. In the chosen basis, where $H^{}_{\rm R} = {\rm diag}\{M^2_1, M^2_2, M^2_3\}$ and $H^{}_{\rm D}$ is given in Eq.~(\ref{eq:HD parameterization}), one can explicitly find
\begin{eqnarray}
\widetilde{\cal I}_3^{\prime} = 6 (M_1^2 - M_2^2) (M_1^2 - M_3^2) (M_2^2 - M_3^2) y^{}_{22} y_{33}^2 y^{}_{21} y^{}_{31} y^{}_{32} \sin \zeta \; ,
\label{eq:Iprime3}
\end{eqnarray}
which is simply proportional to $\sin\zeta$. If the masses of heavy Majorana neutrinos are nondegenerate and the parameters $y^{}_{ij}$ are nonzero, then $\widetilde{\cal I}_3^{\prime} = 0$ is the sufficient and necessary condition for $\zeta = 0$. Now that $\zeta = 0$ is guaranteed by $\widetilde{\cal I}^\prime_3 = 0$, we can calculate the other two invariants, namely,
\begin{eqnarray}
\widetilde{\cal I}^{}_1 & = & M_2^3 \left[ M^{}_3 y_{33}^2 y_{32}^2 + M^{}_1 (y^{}_{22} y^{}_{21} + y^{}_{31} y^{}_{32})^2 + M^{}_2 (y_{22}^2 + y_{32}^2)^2 \right] \sin 2\beta^{}_1 \nonumber \\
&~& + M_3^3 y_{33}^2 \left(M^{}_3 y_{33}^2 + M^{}_1 y_{31}^2 + M^{}_2 y_{32}^2 \right) \sin2 \beta^{}_2 = 0 \; , \\
\widetilde{\cal I}^{}_2 & = & M_2^5 \left[ M^{}_3 y_{33}^2 y_{32}^2 + M^{}_1 (y^{}_{22} y^{}_{21} + y^{}_{31} y^{}_{32})^2 + M^{}_2 (y_{22}^2 + y_{32}^2)^2 \right] \sin 2\beta^{}_1 \nonumber \\
&~& + M_3^5 y_{33}^2 (M^{}_3 y_{33}^2 + M^{}_1 y_{31}^2 + M^{}_2 y_{32}^2) \sin2 \beta^{}_2 = 0 \; .
\end{eqnarray}
The above system of linear homogenous equations of $\sin 2\beta^{}_1$ and $\sin 2\beta^{}_2$ has the unique trivial solutions $\sin 2\beta^{}_1 = 0$ and $\sin 2\beta^{}_2 = 0$, since the determinant of the coefficient matrix is proportional to $(M_3^2 - M_2^2)$ that is nonzero in the case of nondegenerate masses. Therefore, the vanishing of three WB invariants $\{\widetilde{\cal I}^{}_1, \widetilde{\cal I}^{}_2, \widetilde{\cal I}_3^{\prime}\}$ is the sufficient and necessary condition for the vanishing of those three phases in $H^{}_{\rm D}$. After fixing three phases in $H^{}_{\rm D}$, we have another three independent constraints on the remaining phases $\{\xi, \alpha^{}_1, \alpha^{}_2\}$ from Eqs.~(\ref{eq:J4})-(\ref{eq:J6}). However, these equations are in general nonlinear, so there may exist some parameter space, where Eqs.~(\ref{eq:J4})-(\ref{eq:J6}) do not necessarily imply CP conservation, just as shown in Ref.~\cite{Yu:2019ihs} for the effective theory. Without any information about the physical parameters at high-energy scales, such as the heavy Majorana neutrino masses and the matrix elements of $M^{}_{\rm D}$, it is impossible for us to find another set of three invariants to guarantee CP conservation at least in the physically allowed parameter. Therefore, we take Eqs.~(\ref{eq:J4})-(\ref{eq:J6}) as the sufficient and necessary conditions of eliminating the remaining three phases in some particular parameter space.

Although the invariants given in Eqs.~(\ref{eq:J1})-(\ref{eq:J6}) are by construction independent of the flavor basis, it is convenient to calculate them in the special basis where $M^{}_l$ and $M^{}_{\rm R}$ are both diagonal. By inspecting these conditions, we can prove that CP symmetry is conserved if and only if
\begin{eqnarray}
\label{eq:conditions in special basis}
\omega^\alpha_{mn} \equiv \arg \left[\left(M^{}_{\rm D}\right)^{}_{\alpha m}\right] - \arg \left[\left(M^{}_{\rm D}\right)^{}_{\alpha n}\right] = (p^{}_n - p^{}_m)\frac{\pi}{2} + k^{}_\alpha \pi \; ,
\end{eqnarray}
where $p^{}_n$, $p^{}_m$, $k^{}_\alpha$ are arbitrary integers with $m, n = 1, 2, 3$ and $\alpha = e, \mu, \tau$. The above equation gives totally six independent constraints on the phases of $M^{}_{\rm D}$, while the number of independent phases in $M^{}_{\rm D}$ responsible for CP violation is also six. From Eq.~(\ref{eq:conditions in special basis}), we conclude that in the basis where $M^{}_l$ and $M^{}_{\rm R}$ are diagonal, if the masses of heavy Majorana neutrinos are nondegenerate, then the sufficient and necessary conditions for CP conservation are simply that  \emph{(\romannumeral1) the phases of the elements of $M^{}_{\rm D}$ in the same row but different columns can differ only by an integral multiple of $\pi/2$ and (\romannumeral2) the phase differences between two different rows, i.e., $\omega^\alpha_{mn} - \omega^\beta_{mn}$, can only differ by an even multiple of $\pi/2$}.

As a concrete example for the CP violation at high-energy scales, we consider the CP-violating decays of heavy Majorana neutrinos into left-handed lepton and Higgs doublets, i.e., $N^{}_i \to \ell^{}_\alpha + H$ and $N^{}_i \to \overline{\ell}^{}_\alpha + \overline{H}$ (for $i = 1, 2, 3$ and $\alpha = e, \mu, \tau$). The CP asymmetries arise from the interference between the tree and one-loop level decay amplitudes  and can be written as
\begin{eqnarray}
\label{eq:epsilon i alpha}
\epsilon^{}_{i\alpha} \equiv
\frac{\Gamma(N^{}_i \rightarrow \ell^{}_{\alpha} + H) - \Gamma(N^{}_i \rightarrow \overline{\ell}^{}_{\alpha} + \overline{H})}{\sum\limits_{\alpha} \left[ \Gamma(N^{}_i \rightarrow \ell^{}_{\alpha} + H) + \Gamma(N^{}_i \rightarrow \overline{\ell}^{}_{\alpha} + \overline{H}) \right]} \; ,
\end{eqnarray}
where $\Gamma(N^{}_i \to \ell^{}_\alpha + H)$ and $\Gamma(N^{}_i \to \overline{\ell}^{}_\alpha + \overline{H})$ stand for the decay rate of $N^{}_i \to \ell^{}_\alpha + H$ and that of $N^{}_i \to \overline{\ell}^{}_\alpha + \overline{H}$, respectively. In the vanilla scenario of leptogenesis, the CP violation in the out-of-equilibrium decays of heavy Majorana neutrinos gives rise to lepton number asymmetries, which will be finally converted into baryon number asymmetry in our Universe~\cite{Fukugita:1986hr, Davidson:2008bu, Bodeker:2020ghk}. Concentrating on the CP asymmetries, in the basis where $M^{}_l$ and $M^{}_{\rm R}$ are diagonal, we have~\cite{Xing:2011zza}
\begin{eqnarray}
\epsilon^{}_{i\alpha} = \frac{1}{4\pi v^2 \left(H^{}_{\rm D}\right)^{}_{ii}} \sum \limits_{j\neq i} \left\{ {\rm Im} \left[(M_{\rm D}^{*})^{}_{\alpha i}(M^{}_{\rm D})^{}_{\alpha j}(H^{}_{\rm D})^{}_{ij}\right] {\cal F}\left(\frac{M_j^2}{M_i^2}\right) \right. \nonumber \\
+ \left.{\rm Im}\left[(M_{\rm D}^{*})^{}_{\alpha i}(M^{}_{\rm D})^{}_{\alpha j}(H^{}_{\rm D})^{*}_{ij}\right]{\cal G}\left(\frac{M_j^2}{M_i^2}\right) \right\} \; ,
\label{eq:epsilon i alpha expression}
\end{eqnarray}
where the loop functions ${\cal F}(x) \equiv \sqrt{x}\{(2-x)/(1-x) + (1+x)\ln[x/(1+x)]\}$ and ${\cal G}(x) \equiv 1/(1-x)$ have been defined. It is easy to verify that all the CP asymmetries $\epsilon^{}_{i\alpha}$ (for $i = 1, 2, 3$ and $\alpha = e, \mu, \tau$) vanish if the phases of the matrix elements of $M^{}_{\rm D}$ satisfy the following relations
\begin{eqnarray}
\sin\left( \omega^\alpha_{ij} + \omega^\beta_{ij} \right) = \sin \left( \omega^\alpha_{ij} - \omega^\beta_{ij} \right) = 0 \; ,
\label{eq:sinomega}
\end{eqnarray}
where $\omega^\alpha_{ij} \equiv \arg [(M^{}_{\rm D})^{}_{\alpha i}] - \arg [(M^{}_{\rm D})^{}_{\alpha j}]$ has been defined and likewise for $\omega^\beta_{ij}$. The solutions to Eq.~(\ref{eq:sinomega}) are exactly the same as those in Eq.~(\ref{eq:conditions in special basis}). Hence we reach the conclusion that if the phases of the matrix elements of $M^{}_{\rm D}$ fulfill the conditions in Eq.~(\ref{eq:conditions in special basis}) in the basis where $M^{}_l$ and $M^{}_{\rm R}$ are diagonal, then there will be no CP violation in the canonical seesaw model and all the CP asymmetries $\epsilon^{}_{i\alpha}$ vanish in the decays of heavy Majorana neutrinos.

In summary, if the masses of heavy Majorana neutrinos are nondegenerate, we must implement six WB invariants to ensure CP conservation, e.g., those in Eqs.~(\ref{eq:J1})-(\ref{eq:J6}). This conclusion has been obtained in the literature~\cite{Branco:2001pq, Rebelo:2018qsj}. However, if a partial or complete mass degeneracy of heavy Majorana neutrinos is assumed, an immediate question is how many WB invariants we need for CP conservation.

\subsection{Partial mass degeneracy}

If the masses of heavy Majorana neutrinos are partially degenerate, e.g., $M^{}_1 = M^{}_2 \neq M^{}_3$, then one can verify that $\{\widetilde{\cal I}^{}_1, \widetilde{\cal I}^{}_2, \widetilde{\cal I}^{}_3\}$ become linearly dependent on each other, so do $\{ \widetilde{\cal I}^{}_4, \widetilde{\cal I}^{}_5, \widetilde{\cal I}^{}_6\}$. As a consequence, Eqs.~(\ref{eq:J1})-(\ref{eq:J6}) give rise to only two independent equations, which are insufficient to guarantee CP conservation. In this subsection, we attempt to make clear how many CP phases are left in the theory and how to construct the WB invariants for CP conservation in the presence of a partial mass degeneracy.

First, in the basis where $M^{}_l$ and $M^{}_{\rm R}$ are diagonal, we have the freedom to rotate the heavy Majorana neutrino fields as $N^{}_{\rm R} \rightarrow R^{\dagger}_{12}(\alpha) N^{}_{\rm R}$, where $R^{}_{12}(\alpha)$ is the same rotation matrix as given in Sec.~\ref{subsec:partial degeneracy}. Under such a rotation, we have
\begin{eqnarray}
M^{}_{\rm D} \rightarrow M^{}_{\rm D} R^{}_{12}(\alpha), \qquad H^{}_{\rm D} \rightarrow R_{12}^{\dagger}(\alpha) H^{}_{\rm D} R^{}_{12}(\alpha) \; ,
\end{eqnarray}
while the charged-lepton mass matrix $M^{}_l$ and the charged-current interaction are unchanged. Similar to what we have done in Sec.~\ref{subsec:partial degeneracy}, one can adjust $\alpha$ to eliminate one of three CP phases in $H^{}_{\rm D}$, e.g., $\beta^{}_2$. Hence with only two phases left in $H^{}_{\rm D}$, the expression of $\widetilde{\cal I}^{}_1$ becomes quite simple
\begin{eqnarray}
\widetilde{\cal I}^{}_1 = M^{}_1 M^{}_3 (M_1^2 - M_3^2) y_{33}^2 y_{32}^2 \sin \left[ 2(\beta^{}_1 + \zeta)\right] \; .
\label{eq:I1tpart}
\end{eqnarray}
In addition to $\widetilde{\cal I}^{}_1$, inspired by ${\cal I}^{}_3$ in Eq.~(\ref{eq:I3}), we introduce another WB invariant that depends only on the phases in $H^{}_{\rm D}$, namely,
\begin{eqnarray}\label{eq:J7}
\widetilde{\cal I}^{}_7 \equiv {\rm Tr}\left\{\left[G^{}_{\rm DR}, H^{}_{\rm D} \right]^3 \right\}=0 \; .
\end{eqnarray}
After some algebraic calculations, it is easy to verify that $\widetilde{\cal I}^{}_1 = 0$ leads to either $\beta^{}_1 + \zeta = 0$ or $\beta^{}_1 + \zeta = \pi/2$, as indicated by Eq.~(\ref{eq:I1tpart}). Furthermore, we can obtain $\widetilde{\cal I}^{}_7 \propto \sin \beta^{}_1$ in the former case, while $\widetilde{\cal I}^{}_7 \propto \cos \beta^{}_1$ in the latter. Therefore, $\widetilde{\cal I}^{}_1 = 0$ and $\widetilde{\cal I}^{}_7 = 0$ imply either $\beta^{}_1 = \beta^{}_2 = \sigma = 0$ or $\beta^{}_1 = \pi/2, \beta^{}_2 = \zeta =0$, rendering three CP phases $H^{}_{\rm D}$ trivial.

Then, we need another three independent invariants to eliminate the remaining three phases in $M^{}_{\rm D}$. However, as mentioned above, \{$\widetilde{\cal I}^{}_4, \widetilde{\cal I}^{}_5, \widetilde{\cal I}^{}_6$\} turn out to be linearly dependent in the case of a partial mass degeneracy, so they are no longer sufficient to give three independent constraints on the CP phases in $M^{}_{\rm D}$. To this end, we shall construct a new series of WB invariants by using $H^{}_n$ and $G^{}_n$ introduced in Eq.~(\ref{eq:HnGn}). For instance, we introduce
\begin{eqnarray}
\widetilde{\cal I}^{}_8 & \equiv & {\rm Tr}\left\{\left[G^{}_{1}, H^{}_1 \right]^3 \right\} = 0 \; ,\label{eq:J8} \\
\widetilde{\cal I}^{}_9 & \equiv & {\rm Tr}\left\{\left[G^{}_{2}, H^{}_2 \right]^3 \right\} = 0 \; , \label{eq:J9}\\
\widetilde{\cal I}_{10} & \equiv & {\rm Tr}\left\{\left[G^{}_{3}, H^{}_3 \right]^3 \right\} = 0 \; , \label{eq:J10}
\end{eqnarray}
where the explicit expressions of $H^{}_n$ and $G^{}_n$ (for $n = 1, 2, 3$) can be read off from Eq.~(\ref{eq:HnGn}). The construction of three WB invariants in Eqs.~(\ref{eq:J8})-(\ref{eq:J10}) has been motivated by two important observations. First, all the invariants $\{\widetilde{\cal I}^{}_8, \widetilde{\cal I}^{}_9, \widetilde{\cal I}^{}_{10}\}$ are constructed by directly using $M^{}_{\rm D}$ instead of $H^{}_{\rm D}$, so these invariants contain the remaining three CP phases in $M^{}_{\rm D}$. Second, these invariants are similar to each other, but have been constructed intentionally by adopting the charged-lepton mass matrix via $(H^{}_l)^n$ for $n = 1, 2, 3$. In this way, because of the hierarchical mass spectrum of charged leptons, these three invariants are linearly independent even when the masses of heavy Majorana neutrinos are fully degenerate. Therefore, one can constrain the remaining phases in $M^{}_{\rm D}$ to be trivial by requiring $\widetilde{\cal I}^{}_8 = \widetilde{\cal I}^{}_9 = \widetilde{\cal I}^{}_{10} = 0$, whereas those three CP phases in $H^{}_{\rm D}$ have already been eliminated by $\widetilde{\cal I}^{}_1 = \widetilde{\cal I}^{}_7 = 0$.

To conclude, in the presence of a partial mass degeneracy of heavy Majorana neutrinos, the number of CP phases in the theory will be reduced from six to five. In this case, we advocate a new set of WB invariants \{$\widetilde{\cal I}^{}_1, \widetilde{\cal I}^{}_7, \widetilde{\cal I}^{}_8, \widetilde{\cal I}^{}_9, \widetilde{\cal I}^{}_{10}$\}. The vanishing of all these invariants serves as the sufficient and necessary condition for CP conservation in this particular case.

\subsection{Complete mass degeneracy}

Once the masses of heavy Majorana neutrinos are completely degenerate, i.e., $M^{}_1 = M^{}_2 = M^{}_3$, all the six WB invariants in Eqs.~(\ref{eq:J1})-(\ref{eq:J6}) will automatically vanish. Therefore, they will not carry any useful information about CP violation.

In the presence of full mass degeneracy, however, we are allowed to perform an arbitrary orthogonal rotation of $M^{}_{\rm R}$ in the basis where both $M^{}_l$ and $M^{}_{\rm R}$ are diagonal, without changing the heavy Majorana neutrino mass term. As we have proved in Sec.~\ref{subsec:complete degeneracy}, these three degrees of freedom in the arbitrary orthogonal rotation can be taken to reduce the number CP phases in $H^{}_{\rm D}$ at most by two, so we are left with four CP phases in total.

It is obvious that these four CP phases can be made trivial by requiring four WB invariants in Eq.~(\ref{eq:J7}) and in Eqs.~(\ref{eq:J8})-(\ref{eq:J10}) to be zero. First, $\widetilde{\cal I}^{}_7 = 0$ can be used to get rid of the only CP phase in $H^{}_{\rm D}$, as the other two phases have been removed by two successive rotations. Then, the vanishing of $\{\widetilde{\cal I}^{}_8, \widetilde{\cal I}^{}_9, \widetilde{\cal I}^{}_{10} \}$ in Eqs.~(\ref{eq:J8})-(\ref{eq:J10}) guarantees that three remaining phases in $M^{}_{\rm D}$ are trivial. Therefore, for the complete mass degeneracy of heavy Majorana neutrinos, there are four CP phases and the vanishing of the WB invariants \{$\widetilde{\cal I}^{}_7, \widetilde{\cal I}^{}_8, \widetilde{\cal I}^{}_9, \widetilde{\cal I}^{}_{10}$\} serves as the sufficient and necessary condition for CP conservation.

It is worth stressing that the partial or complete mass degeneracy of heavy Majorana neutrinos may be guaranteed by flavor symmetries or simply accidental, and thus the degeneracy will be shifted by explicit symmetry breaking or radiative corrections~\cite{GonzalezFelipe:2003fi, Turzynski:2004xy, Joaquim:2005zv, Branco:2005ye, Dev:2014laa, Pilaftsis:2015bja}, leading to the possibility of successful resonant leptogenesis~\cite{Pilaftsis:2003gt, Pilaftsis:2005rv}. Moreover, once the mass degeneracy is broken, the number of CP-violating phases and relevant WB invariants will be changed, as discussed in Sec.~\ref{subsec:complete theory nondegerate mass}.

In the presence of either partial or complete mass degeneracy of heavy Majorana neutrinos, the CP asymmetries defined in Eq.~(\ref{eq:epsilon i alpha}) cannot be simply obtained from Eq.~(\ref{eq:epsilon i alpha expression}), which turns out to be singular in the exact degeneracy limit (i.e., $M^{}_{i} = M^{}_{j}$). When the resonant mixing between any two nearly-degenerate unstable particles is properly treated~\cite{Pilaftsis:1997jf, Pilaftsis:2003gt, Pilaftsis:2005rv, Anisimov:2005hr, Dev:2017wwc}, the divergence arising from one-loop self-energy corrections to the heavy Majorana neutrino decays can be removed. After taking account of both one-loop self-energy and vertex corrections, one can find that the loop functions in the expressions of CP asymmetries in Eq.~(\ref{eq:epsilon i alpha expression}) are modified with a regulator~\cite{Zhang:2015tea},
\begin{eqnarray}
\label{eq:loop function mod}
{\cal F}(x_{ij})&=&\sqrt{x_{ij}^{}}\left[\frac{1-x_{ij}^{}}{(1-x_{ij}^{})^2+r_{ij}^2}+1+(1+x_{ij}^{})\,{\rm ln}\,\left(\frac{x_{ij}^{}}{1+x_{ij}^{}}\right) \right]\;,\nonumber \\
 {\cal G}(x_{ij}^{})&=&\frac{1-x_{ij}^{}}{(1-x_{ij}^{})^2+r_{ij}^{2}}\;,
\end{eqnarray}
where $x^{}_{ij} \equiv M_i^2/M_j^2$ has been defined and the regulator $r^{}_{ij}$ has been introduced~\cite{Pilaftsis:2003gt, Pilaftsis:2005rv, Anisimov:2005hr, Dev:2017wwc}. If the mass spectrum of heavy Majorana neutrinos is strongly hierarchical, then the loop functions defined in Eq.~(\ref{eq:loop function mod}) will be reduced to the forms below Eq.~(\ref{eq:epsilon i alpha expression}). However, when the masses of heavy Majorana neutrinos become nearly degenerate, the regulator $r^{}_{ij}$ will play an important role. In particular, in the limit of exact mass degeneracy, i.e., $x_{ij}^{}=1$, the regulator removes the singularity and gives a physically meaningful result.

It deserves to emphasize that in the limit of complete mass degeneracy (i.e., $M_1^{} = M_2^{} = M_3^{}$), although the CP asymmetries $\epsilon^{}_{i\alpha}$ defined in Eq.~(\ref{eq:epsilon i alpha}) remain nonvanishing due to the contribution from the interference between the tree-level amplitude and the one-loop vertex correction, there are actually no CP asymmetries in the decays of heavy Majorana neutrinos~\cite{Pilaftsis:1997jf, Pilaftsis:2003gt}. This is because the CP-violating source terms contributing to the generation of lepton number asymmetry (i.e., the difference between the number density of leptons and that of antileptons) in the Boltzmann equations depend only on the following combinations of CP asymmetries from different generations of heavy Majorana neutrinos in the limit of complete mass degeneracy~\cite{Pilaftsis:2003gt}, namely,
\begin{eqnarray}
\label{eq:epsilon effective}
\epsilon^{}_{\rm{eff}}\equiv \frac{\sum\limits_{i=1}^{3} \left[\Gamma(N^{}_i \rightarrow \ell^{}_{\alpha} + H) - \Gamma(N^{}_i \rightarrow \overline{\ell}^{}_{\alpha} + \overline{H})\right]}{\sum\limits_{i=1}^{3}\sum\limits_{\alpha} \left[ \Gamma(N^{}_i \rightarrow \ell^{}_{\alpha} + H) + \Gamma(N^{}_i \rightarrow \overline{\ell}^{}_{\alpha} + \overline{H}) \right]} \; ,
\end{eqnarray}
which turn out to be vanishing. This can be understood by noticing that all the heavy Majorana neutrinos in the mass-degeneracy limit contribute to the generation of CP asymmetries and only the effective CP asymmetries defined in Eq.~(\ref{eq:epsilon effective}) play a role in leptogenesis.

To conclude, although there remain four nonvanishing CP phases in the limit of complete mass degeneracy, there are actually no CP asymmetries in the decays of heavy Majorana neutrinos.

\subsection{Minimal seesaw model}

In this subsection, we examine the so-called minimal seesaw model (MSM), in which only two right-handed neutrino singlets are introduced~\cite{Kleppe:1995zz, Ma:1998zg, King:1999mb, King:2002nf, Frampton:2002qc}. See, e.g., Refs.~\cite{Guo:2006qa, Xing:2020ald}, for recent reviews on the MSM. In this minimal scenario, $M^{}_{\rm D}$ is actually a $3\times 2$ complex matrix, and the effective mass matrix of three light Majorana neutrinos is given by the seesaw formula $M^{}_\nu = - M^{}_{\rm D} M^{-1}_{\rm R} M^{\rm T}_{\rm D}$. As is well known, the rank of $M^{}_\nu$ will thus be at most two, indicating that the lightest neutrino is massless. Without loss of generality, we take $m^{}_1 = 0$ for the normal mass ordering for illustration.

Although $M^{}_{\rm D}$ generally contains six phases, three of them are actually unphysical and can be removed by the basis transformations of lepton fields $\nu^{}_{\rm L}$, $l^{}_{\rm L}$ and $l^{}_{\rm R}$. In the following discussions, we take the Casas-Ibarra parametrization of $M^{}_{\rm D}$~\cite{Casas:2001sr, Ibarra:2003up}, i.e.,
\begin{eqnarray}
M^{}_{\rm D} = {\rm i} U \sqrt{\widehat{M}^{}_{\nu}} R \sqrt{\widehat{M}^{}_{\rm R}} \; ,
\end{eqnarray}
where the PMNS matrix $U$ can be decomposed as $U = V \cdot \text{diag}\left\{1, e^{{\rm i} \sigma}, 1\right\}$,\footnote{As the lightest neutrino is massless, the Majorana CP phase associated with the corresponding neutrino mass eigenstate disappears from the theory.} with $V$ being the CKM-like matrix that contains one Dirac CP phase $\delta$ and three mixing angles. In addition, both light and heavy Majorana neutrino mass matrices $\widehat{M}^{}_{\nu} = \text{diag}\{0, m^{}_2, m^{}_3\}$ and  $\widehat{M}^{}_{\rm R} = \text{diag}\{M^{}_1, M^{}_2\}$ are diagonal, and the complex and orthogonal matrix $R$, satisfying $R^{\rm T} R = \text{diag}\{1, 1\}$ and $R R^{\rm T} = {\rm diag}\{0, 1, 1\}$, can be parameterized as~\cite{Ibarra:2003up}
\begin{eqnarray}
R = \left(
\begin{array}{cc}
0&0\\
\cos z&-\sin z\\
\pm \sin z&\pm \cos z\\
\end{array}
\right) \; ,
\end{eqnarray}
where $z$ is an arbitrary complex number. With such a parametrization, one can observe that one CP phase of $M^{}_{\rm D}$ is located in $R$, while the other two are included in the PMNS matrix $U$.

Now we explain how to construct the WB invariants in the MSM and present the sufficient and necessary conditions for CP conservation in the cases of nondegenerate (i.e., $M^{}_1\neq M^{}_2$) and degenerate (i.e., $M^{}_1 = M^{}_2$) heavy Majorana neutrino masses.
\begin{itemize}
\item For $M^{}_1 \neq M^{}_2$, there are totally three CP phases in $M^{}_{\rm D}$, for which one has to construct three WB invariants to guarantee CP conservation. In the MSM, however, only two out of those six invariants in Eqs.~(\ref{eq:J1})-(\ref{eq:J6}) are linearly independent, and we choose $\widetilde{\cal I}^{}_1$ and $\widetilde{\cal I}^{}_4$. As one can see from the definition of $\widetilde{\cal I}^{}_1$ in Eq.~(\ref{eq:J1}), only $H^{}_{\rm D}$ is involved in this invariant, so it contains the unique CP phase in $R$. On the other hand, $\widetilde{\cal I}^{}_4$ defined in Eq.~(\ref{eq:J4}) depend on the CP phase in $R$ as well as two CP phases in the PMNS matrix $U$. For this reason, we need to construct extra WB invariants, in which the CP phases in $U$ are present. Unfortunately, all the invariants $\{\widetilde{\cal I}^{}_7, \widetilde{\cal I}^{}_8, \widetilde{\cal I}^{}_9, \widetilde{\cal I}^{}_{10}\}$ in Eqs.~(\ref{eq:J7})-(\ref{eq:J10}) vanish automatically in the MSM.

    Inspired by the invariants $\{{\cal I}^{}_1, {\cal I}^{}_2, {\cal I}^{}_3, {\cal I}^{}_4\}$ in the effective theory, we can simply replace $M^{}_\nu$ by $-M^{}_{\rm D} M^{-1}_{\rm R} M^{\rm T}_{\rm D}$ everywhere in these invariants and then obtain four nontrivial WB invariants in the MSM, i.e.,
\begin{eqnarray}
\widehat{\cal I}^{}_{1} &\equiv& {\rm Tr}\left\{ \left[ M^{}_{\rm D} M_{\rm R}^{-1} H^*_{\rm D} (M_{\rm R}^{-1})^{\dagger} M_{\rm D}^{\dagger}, H^{}_l \right]^3 \right \} \; ,
\label{eq:J11}\\
\widehat{\cal I}^{}_{2} &\equiv& {\rm Im}\left\{ {\rm Tr}\left[ H^{}_l M^{}_{\rm D} M_{\rm R}^{-1} H_{\rm D}^{*} (M_{\rm R}^{-1})^{\dagger} H^{}_{\rm D} M_{\rm R}^{-1} M_{\rm D}^{T} H_l^{*}  M_{\rm D}^{*} (M_{\rm R}^{-1})^{\dagger} M_{\rm D}^{\dagger}\right]\right\} \; ,
\label{eq:J12}\\
\widehat{\cal I}^{}_{3} &\equiv& {\rm Tr}\left\{ \left[M^{}_{\rm D} M_{\rm R}^{-1} M_{\rm D}^{\rm T} H_l^{*}  M_{\rm D}^{*} (M_{\rm R}^{-1})^{\dagger} M_{\rm D}^{\dagger}, H^{}_l \right]^3 \right\} \; ,
\label{eq:J13}\\
\widehat{\cal I}^{}_{4} &\equiv& {\rm Im}\left\{ {\rm Tr}\left[H^{}_l (M^{}_{\rm D} M_{\rm R}^{-1} H_{\rm D}^{*} (M_{\rm R}^{-1})^{\dagger} M_{\rm D}^{\dagger})^{2} M^{}_{\rm D} M_{\rm R}^{-1} M_{\rm D}^{\rm T} H_l^{*}  M_{\rm D}^{*} (M_{\rm R}^{-1})^{\dagger} M_{\rm D}^{\dagger}  \right] \right\} \; . \label{eq:J14}
\end{eqnarray}
It should be noted that $\widehat{\cal I}^{}_i$ (for $i = 1, 2, 3, 4$) depend only on two CP phases in the PMNS matrix $U$ and have nothing to do with the CP phase in $R$. Moreover, $\widehat{\cal I}^{}_{1}$ is proportional to $\sin\delta$, where $\delta$ is the Dirac-type CP phase in the PMNS matrix $U$, but not related to the Majorana-type CP phase $\sigma$. In contrast, $\{\widehat{\cal I}_{2}, \widehat{\cal I}_{3}, \widehat{\cal I}_{4}\}$ depend on both $\delta$ and $\sigma$. With all these invariants, to guarantee CP conservation, we can first require $\widetilde{\cal I}^{}_1 = 0$ to render the phase in $R$ trivial, then $\widehat{\cal I}^{}_{1} = 0$ to eliminate $\delta$ in $U$, and finally either $\widetilde{\cal I}^{}_4 = 0$ or one of $\{\widehat{\cal I}_{2} = 0, \widehat{\cal I}_{3} = 0, \widehat{\cal I}_{4} = 0\}$ to get rid of $\sigma$ in $U$.

\item For $M^{}_1 = M^{}_2$, similar to the case of partial mass degeneracy in the effective theory or in the canonical seesaw model,  there is an extra degree of freedom in the system, which can be implemented to remove the only CP phase in $R$. Therefore, we are left with two CP phases. It is straightforward to verify that $\widetilde{\cal I}^{}_1$ and $\widetilde{\cal I}^{}_4$ vanish automatically in this limit of $M^{}_1 = M^{}_2$. However, since $\{\widehat{\cal I}_{1}, \widehat{\cal I}_{2}, \widehat{\cal I}_{3}, \widehat{\cal I}_{4}\}$ are independent of heavy Majorana neutrino masses, they are in general nonzero in the presence of mass degeneracy. We can first use $\widehat{\cal I}^{}_{1} = 0$ to make $\delta$ in $U$ trivial, and then choose any one of $\{\widehat{\cal I}_{2} = 0, \widehat{\cal I}_{3} = 0, \widehat{\cal I}_{4} = 0\}$ to eliminate the remaining phase $\sigma$ in $U$, so that CP conservation is guaranteed.
\end{itemize}
In summary, if there is no mass degeneracy of heavy Majorana neutrinos, we have three CP phases and the vanishing of three WB invariants $\{\widetilde{\cal I}^{}_1, \widetilde{\cal I}^{}_4, \widehat{\cal I}^{}_1\}$ serves as the sufficient and necessary condition for CP conservation. In addition, in the case of mass degeneracy, there are two CP phases and one can find that CP conservation is ensured by $\{\widehat{\cal I}^{}_1 = 0, \widehat{\cal I}^{}_2 = 0\}$. It is worth mentioning that the choice of WB invariants is by no means unique, but different choices are all equivalent.

\section{Summary}\label{sec:summary}
\begin{table}[t!]
\centering
\begin{tabular}{l|c|l}
\hline \hline
{\bf Low-energy Effective Theory} & Number of CP phases & Weak-Basis Invariants \\
\hline
\hline
\multirow{3}{*}{No degeneracy ($m^{}_1 \neq m^{}_2 \neq m^{}_3$)} & \multirow{3}{*}{3} & ${\cal I}^{}_1 \equiv {\rm Tr}\left\{ \left[H^{}_\nu, H^{}_l \right]^3\right\}$ \\
~ & ~ & ${\cal I}^{}_2 \equiv {\rm Im}\left\{{\rm Tr}\left[H^{}_l H^{}_\nu G^{}_{l\nu}\right]\right\}$ \\
~ & ~ & ${\cal I}_4 \equiv {\rm Im}\left\{{\rm Tr}\left[H^{}_l H^{2}_\nu G^{}_{l\nu}\right]\right\}$ \\
\hline
\multirow{2}{*}{Partial degeneracy ($m^{}_1 = m^{}_2 \neq m^{}_3$)} & \multirow{2}{*}{2} & ${\cal I}^{}_2 \equiv {\rm Im}\left\{{\rm Tr}\left[H^{}_l H^{}_\nu G^{}_{l\nu}\right]\right\} $ \\
~ & ~ & ${\cal I}^{}_3 \equiv {\rm Tr}\left\{ \left[G^{}_{l\nu}, H^{}_l \right]^3\right\}$ \\
\hline
Full degeneracy ($m^{}_1 = m^{}_2 = m^{}_3$) & 1 & ${\cal I}^{}_3 \equiv {\rm Tr}\left\{ \left[G^{}_{l\nu}, H^{}_l \right]^3\right\}$ \\
\hline
\multirow{2}{*}{No degeneracy with $m^{}_1 = 0$} & \multirow{2}{*}{2} &  ${\cal I}^{}_1 \equiv {\rm Tr}\left\{ \left[H^{}_\nu, H^{}_l \right]^3\right\}$ \\
~ & ~ & ${\cal I}^{}_2 \equiv {\rm Im}\left\{{\rm Tr}\left[H^{}_l H^{}_\nu G^{}_{l\nu}\right]\right\}$ \\
\hline \hline
\end{tabular}
\vspace{0.2cm}
\caption{Summary of the number of independent CP phases and the weak-basis invariants chosen to guarantee CP conservation in the low-energy effective theory. Notice that the choice of weak-basis invariants is by no means unique.}
\label{table:effective theory}
\end{table}

\renewcommand\arraystretch{1.19}
\begin{table}[t!]
\centering
\begin{tabular}{l|c|l}
\hline \hline
{\bf Canonical Seesaw Model} & Number of CP phases & Weak-Basis Invariants \\
\hline
\hline
\multirow{6}{*}{No degeneracy ($M^{}_1 \neq M^{}_2 \neq M^{}_3$)} & \multirow{6}{*}{6} & $\widetilde{\cal I}^{}_1 \equiv {\rm Im}\left\{{\rm Tr}\left[H^{}_{\rm D} H^{}_{\rm R} G^{}_{\rm DR}\right]\right\}$ \\
~ & ~ & $\widetilde{\cal I}^{}_2 \equiv {\rm Im}\left\{{\rm Tr}\left[H^{}_{\rm D} H^{2}_{\rm R} G^{}_{\rm DR}\right]\right\}$ \\
~ & ~ & $\widetilde{\cal I}_3^{\prime} \equiv {\rm Tr} \left\{ \left[ H^{}_{\rm R}, H^{}_{\rm D} \right]^3 \right\}$ \\
~ & ~ & $\widetilde{\cal I}^{}_4 \equiv {\rm Im}\left\{{\rm Tr}\left[H^{}_{1} H^{}_{\rm R} G^{}_{1}\right]\right\}$ \\
~ & ~ & $\widetilde{\cal I}^{}_5 \equiv {\rm Im}\left\{{\rm Tr}\left[H^{}_{1} H^2_{\rm R} G^{}_{1}\right]\right\}$ \\
~ & ~ & $\widetilde{\cal I}^{}_6 \equiv {\rm Im}\left\{{\rm Tr}\left[H^{}_{1} H^{}_{\rm R} G^{}_{1}H^{}_{\rm R}\right]\right\}$ \\
\hline
\multirow{5}{*}{Partial degeneracy ($M^{}_1 = M^{}_2 \neq M^{}_3$)} & \multirow{5}{*}{5} & $\widetilde{\cal I}^{}_1 \equiv {\rm Im}\left\{{\rm Tr}\left[H^{}_{\rm D} H^{}_{\rm R} G^{}_{\rm DR}\right]\right\}$ \\
~ & ~ & $\widetilde{\cal I}^{}_7 \equiv {\rm Tr}\left\{\left[G^{}_{\rm DR}, H^{}_{\rm D} \right]^3 \right\}$ \\
~ & ~ & $\widetilde{\cal I}^{}_8 \equiv {\rm Tr}\left\{\left[G^{}_{1}, H^{}_1 \right]^3 \right\}$ \\
~ & ~ & $\widetilde{\cal I}^{}_9 \equiv {\rm Tr}\left\{\left[G^{}_{2}, H^{}_2 \right]^3 \right\}$ \\
~ & ~ & $\widetilde{\cal I}^{}_{10} \equiv {\rm Tr}\left\{\left[G^{}_{3}, H^{}_3 \right]^3 \right\}$ \\
\hline
\multirow{4}{*}{Full degeneracy ($M^{}_1 = M^{}_2 = M^{}_3$)} & \multirow{4}{*}{4} & $\widetilde{\cal I}^{}_7 \equiv {\rm Tr}\left\{\left[G^{}_{\rm DR}, H^{}_{\rm D} \right]^3 \right\}$ \\
~ & ~ & $\widetilde{\cal I}^{}_8 \equiv {\rm Tr}\left\{\left[G^{}_{1}, H^{}_1 \right]^3 \right\}$ \\
~ & ~ & $\widetilde{\cal I}^{}_9 \equiv {\rm Tr}\left\{\left[G^{}_{2}, H^{}_2 \right]^3 \right\}$ \\
~ & ~ & $\widetilde{\cal I}^{}_{10} \equiv {\rm Tr}\left\{\left[G^{}_{3}, H^{}_3 \right]^3 \right\}$ \\
\hline
\multirow{3}{*}{Minimal seesaw model ($M^{}_1 \neq M^{}_2$)} & \multirow{3}{*}{3} & $\widetilde{\cal I}^{}_1 \equiv {\rm Im}\left\{{\rm Tr}\left[H^{}_{\rm D} H^{}_{\rm R} G^{}_{\rm DR}\right]\right\}$ \\
~ & ~ & $\widetilde{\cal I}^{}_4 \equiv {\rm Im}\left\{{\rm Tr}\left[H^{}_{1} H^{}_{\rm R} G^{}_{1}\right]\right\}$ \\
~ & ~ & $\widehat{\cal I}^{}_1$ in Eq.~(\ref{eq:J11}) \\
\hline
\multirow{2}{*}{Minimal seesaw model ($M^{}_1 = M^{}_2$)} & \multirow{2}{*}{2} &  $\widehat{\cal I}^{}_1$ in Eq.~(\ref{eq:J11}) \\
~ & ~ & $\widehat{\cal I}^{}_2$ in Eq.~(\ref{eq:J12}) \\
\hline \hline
\end{tabular}
\vspace{0.3cm}
\caption{Summary of the number of independent CP phases and the weak-basis invariants chosen to guarantee CP conservation in the canonical seesaw model. Notice that the choice of weak-basis invariants is by no means unique.}
\label{table:complete theory}
\end{table}
\renewcommand\arraystretch{1}
In this paper, we have performed a systematic study of the sufficient and necessary conditions for CP conservation in leptonic sector, both in the low-energy effective theory of massive Majorana neutrinos and in the canonical seesaw model. A particular attention has been paid to the cases of the mass degeneracy of either light or heavy Majorana neutrinos. We have demonstrated how to count correctly the number of independent CP phases in these cases, and explained the strategy to construct the WB invariants to guarantee CP conservation.

In the low-energy effective theory, if the masses of light Majorana neutrinos are not degenerate, there are totally three independent CP phases. If the masses of light neutrinos are partially or completely degenerate, then there will be extra degrees of freedom in the theory allowing us to rotate the left-handed neutrino fields without changing their mass term. As a consequence, such degrees of freedom can be used to reduce the number of independent CP phases. The number of CP phases and the WB invariants chosen to guarantee CP conservation in different cases are summarized in Table~\ref{table:effective theory}. Moreover, the renormalization-group equations of the WB invariants in the effective theory have been derived. By using these equations of WB invariants, we show that CP conservation will not be violated by radiative corrections.

In the canonical seesaw model, there are totally six independent CP phases in the case of nondegenerate masses of heavy Majorana neutrinos. Just like in the effective theory, in the presence of mass degeneracy, it is possible to reduce the number of CP phases. The main results have been summarized in Table~\ref{table:complete theory}. The sufficient and necessary conditions for CP conservation in the minimal seesaw model are also given. In the basis where the charged-lepton mass matrix $M^{}_l$ and right-handed neutrino mass matrix $M^{}_{\rm R}$ are diagonal, the conserved CP symmetry would lead to
the vanishing of all flavor-dependent CP asymmetries in the heavy Majorana neutrino decays, i.e., $\epsilon^{}_{i\alpha}$ for $i = 1, 2, 3$ and $\alpha = e, \mu, \tau$, while any nonzero CP asymmetries imply the existence of CP violation. It is worth pointing out that a flavor symmetry must be introduced to protect the mass degeneracy. Otherwise, either partial or complete mass degeneracy of heavy Majorana neutrinos will be violated by radiative corrections~\cite{GonzalezFelipe:2003fi, Turzynski:2004xy, Joaquim:2005zv, Branco:2005ye, Dev:2014laa, Pilaftsis:2015bja} and all those six WB invariants in Eqs.~(\ref{eq:J1})-(\ref{eq:J6}) are needed to guarantee CP conservation.

We stress that the choice of different sets of WB invariants for CP conservation is not unique. In each case, we have explicitly given a suitable set of WB invariants, which should be useful for the future studies of leptonic CP violation and for the model building of neutrino mass generation and lepton flavor mixing.

\section*{Acknowledgements}

The authors thank Prof. Thomas Schwetz and Prof. Zhi-zhong Xing for helpful discussions. This work was supported in part by the National Natural Science Foundation of China under grant No.~11775232 and No.~11835013, and by the CAS Center for Excellence in Particle Physics.


\begin{thebibliography}{99}



\bibitem{Tanabashi:2018oca}
  M.~Tanabashi {\it et al.} [ParticleDataGroup],
  ``Review of Particle Physics,''
  Phys.\ Rev.\ D {\bf 98}, no. 3, 030001 (2018).

\bibitem{Xing:2019vks}
  Z.~z.~Xing,
  ``Flavor structures of charged fermions and massive neutrinos,''
  Phys.\ Rept.\  {\bf 854}, 1 (2020)
  [arXiv:1909.09610].

\bibitem{Branco:2011zb}
  G.~C.~Branco, R.~G.~Felipe and F.~R.~Joaquim,
  ``Leptonic CP Violation,''
  Rev.\ Mod.\ Phys.\  {\bf 84}, 515 (2012)
  [arXiv:1111.5332].

\bibitem{Majorana:1937vz}
  E.~Majorana,
  ``Teoria simmetrica dell' elettrone e del positrone,''
  Nuovo Cim.\  {\bf 14}, 171 (1937).

\bibitem{Racah:1937qq}
  G.~Racah,
  ``On the symmetry of particle and antiparticle,''
  Nuovo Cim.\  {\bf 14}, 322 (1937).


\bibitem{Minkowski}
  P.~Minkowski,
  ``$\mu \to e\gamma$ at a Rate of One Out of $10^{9}$ Muon Decays?,''
  Phys.\ Lett.\  {\bf 67B}, 421 (1977).

\bibitem{Yanagida}
  T.~Yanagida,
  ``Horizontal Symmetry And Masses Of Neutrinos,''
  Conf.\ Proc.\ C {\bf 7902131}, 95 (1979).

\bibitem{Gell-Mann}
  M.~Gell-Mann, P.~Ramond and R.~Slansky,
  ``Complex Spinors and Unified Theories,''  Conf.\ Proc.\ C {\bf 790927}, 315 (1979)  [arXiv:1306.4669].

\bibitem{Glashow}
  S.~L.~Glashow,
  ``The Future of Elementary Particle Physics,''
  NATO Sci.\ Ser.\ B {\bf 61}, 687 (1980).

\bibitem{Mohapatra}
  R.~N.~Mohapatra and G.~Senjanovic,
  ``Neutrino Mass and Spontaneous Parity Violation,''
  Phys.\ Rev.\ Lett.\  {\bf 44}, 912 (1980).

\bibitem{Pontecorvo:1957cp}
  B.~Pontecorvo,
  ``Mesonium and anti-mesonium,''
  Sov.\ Phys.\ JETP {\bf 6}, 429 (1957)
  [Zh.\ Eksp.\ Teor.\ Fiz.\  {\bf 33}, 549 (1957)].

\bibitem{Maki:1962mu}
  Z.~Maki, M.~Nakagawa and S.~Sakata,
  ``Remarks on the unified model of elementary particles,''
  Prog.\ Theor.\ Phys.\  {\bf 28}, 870 (1962).

\bibitem{Sakharov:1967dj}
  A.~D.~Sakharov,
  ``Violation of CP Invariance, C asymmetry, and baryon asymmetry of the universe,''
  Pisma Zh.\ Eksp.\ Teor.\ Fiz.\  {\bf 5}, 32 (1967)
  [JETP Lett.\  {\bf 5}, 24 (1967)].

\bibitem{Fukugita:1986hr}
  M.~Fukugita and T.~Yanagida,
  ``Baryogenesis Without Grand Unification,''
  Phys.\ Lett.\ B {\bf 174}, 45 (1986).

\bibitem{Davidson:2008bu}
  S.~Davidson, E.~Nardi and Y.~Nir,
  ``Leptogenesis,''
  Phys.\ Rept.\  {\bf 466}, 105 (2008)
  [arXiv:0802.2962].

\bibitem{Bodeker:2020ghk}
  D.~Bodeker and W.~Buchmuller,
  ``Baryogenesis from the weak scale to the GUT scale,''
  arXiv:2009.07294.

\bibitem{Branco:1986gr}
  G.~C.~Branco, L.~Lavoura and M.~N.~Rebelo,
  ``Majorana Neutrinos and {CP} Violation in the Leptonic Sector,''
  Phys.\ Lett.\ B {\bf 180}, 264 (1986).


\bibitem{Bernabeu:1986fc}
  J.~Bernabeu, G.~C.~Branco and M.~Gronau,
  ``CP Restrictions on Quark Mass Matrices,''
  Phys.\ Lett.\  {\bf 169B}, 243 (1986).

\bibitem{Pilaftsis:1997jf}
A.~Pilaftsis,
``CP violation and baryogenesis due to heavy Majorana neutrinos,''
Phys. Rev. D \textbf{56}, 5431-5451 (1997)
[arXiv:hep-ph/9707235].


\bibitem{Branco:2001pq}
  G.~C.~Branco, T.~Morozumi, B.~M.~Nobre and M.~N.~Rebelo,
  ``A Bridge between CP violation at low-energies and leptogenesis,''
  Nucl.\ Phys.\ B {\bf 617}, 475 (2001)
  [hep-ph/0107164].

\bibitem{Cirigliano:2006nu}
  V.~Cirigliano, G.~Isidori and V.~Porretti,
  ``CP violation and Leptogenesis in models with Minimal Lepton Flavour Violation,''
  Nucl.\ Phys.\ B {\bf 763}, 228 (2007)
  [hep-ph/0607068].

\bibitem{Dreiner:2007yz}
  H.~K.~Dreiner, J.~S.~Kim, O.~Lebedev and M.~Thormeier,
  ``Supersymmetric Jarlskog invariants: The Neutrino sector,''
  Phys.\ Rev.\ D {\bf 76}, 015006 (2007)
  [hep-ph/0703074].

\bibitem{Yu:2019ihs}
  B.~Yu and S.~Zhou,
  ``The number of sufficient and necessary conditions for CP conservation with Majorana neutrinos: three or four?,''
  Phys.\ Lett.\ B {\bf 800}, 135085 (2020)
  [hep-ph/1908.09306].

\bibitem{Branco:1998bw}
  G.~C.~Branco, M.~N.~Rebelo and J.~I.~Silva-Marcos,
  ``Degenerate and quasidegenerate Majorana neutrinos,''
  Phys.\ Rev.\ Lett.\  {\bf 82}, 683 (1999)
  [hep-ph/9810328].

\bibitem{Mei:2003gu}
  J.~w.~Mei and Z.~z.~Xing,
  ``Impact of fermion mass degeneracy on flavor mixing,''
  J.\ Phys.\ G {\bf 30}, 1243 (2004)
  [hep-ph/0312382].



\bibitem{Rebelo:2018qsj}
  M.~N.~Rebelo,
  ``Neutrino Physics and Leptonic Weak Basis Invariants,''
  arXiv:1804.05777.

\bibitem{Jarlskog:1985ht}
  C.~Jarlskog,
  ``Commutator of the Quark Mass Matrices in the Standard Electroweak Model and a Measure of Maximal CP Violation,''
  Phys.\ Rev.\ Lett.\  {\bf 55}, 1039 (1985).

\bibitem{Wu:1985ea}
  D.~d.~Wu,
  ``The Rephasing Invariants and CP,''
  Phys.\ Rev.\ D {\bf 33}, 860 (1986).

\bibitem{Cheng:1986in}
  H.~Y.~Cheng,
  ``{Kobayashi-Maskawa} Type of Hard {CP} Violation Model With Three Generation Majorana Neutrinos,''
  Phys.\ Rev.\ D {\bf 34}, 2794 (1986).

\bibitem{Feldmann:2015nia}
  T.~Feldmann, T.~Mannel and S.~Schwertfeger,
  ``Renormalization Group Evolution of Flavour Invariants,''
  JHEP {\bf 1510}, 007 (2015)
  [arXiv:1507.00328].

\bibitem{Chankowski:1993tx}
  P.~H.~Chankowski and Z.~Pluciennik,
 ``Renormalization group equations for seesaw neutrino masses,''
  Phys.\ Lett.\ B {\bf 316}, 312 (1993)
  [hep-ph/9306333].

\bibitem{Babu:1993qv}
  K.~S.~Babu, C.~N.~Leung and J.~T.~Pantaleone,
  ``Renormalization of the neutrino mass operator,''
  Phys.\ Lett.\ B {\bf 319}, 191 (1993)
  [hep-ph/9309223].

\bibitem{Antusch:2001ck}
  S.~Antusch, M.~Drees, J.~Kersten, M.~Lindner and M.~Ratz,
  ``Neutrino mass operator renormalization revisited,''
  Phys.\ Lett.\ B {\bf 519}, 238 (2001)
  [hep-ph/0108005].

\bibitem{Antusch:2002rr}
  S.~Antusch, J.~Kersten, M.~Lindner and M.~Ratz,
  ``Neutrino mass matrix running for nondegenerate seesaw scales,''
  Phys.\ Lett.\ B {\bf 538}, 87 (2002)
  [hep-ph/0203233].

\bibitem{Antusch:2003kp}
  S.~Antusch, J.~Kersten, M.~Lindner and M.~Ratz,
  ``Running neutrino masses, mixings and CP phases: Analytical results and phenomenological consequences,''
  Nucl.\ Phys.\ B {\bf 674}, 401 (2003)
  [hep-ph/0305273].

\bibitem{Mei:2003gn}
  J.~w.~Mei and Z.~z.~Xing,
  ``Radiative corrections to neutrino mixing and CP violation in the minimal seesaw model with leptogenesis,''
  Phys.\ Rev.\ D {\bf 69}, 073003 (2004)
  [hep-ph/0312167].

\bibitem{Antusch:2005gp}
  S.~Antusch, J.~Kersten, M.~Lindner, M.~Ratz and M.~A.~Schmidt,
  ``Running neutrino mass parameters in see-saw scenarios,''
  JHEP {\bf 0503}, 024 (2005)
  [hep-ph/0501272].

\bibitem{Mei:2005qp}
  J.~w.~Mei,
  ``Running neutrino masses, leptonic mixing angles and CP-violating phases: From M(Z) to Lambda(GUT),''
  Phys.\ Rev.\ D {\bf 71}, 073012 (2005)
  [hep-ph/0502015].

\bibitem{Luo:2005sq}
  S.~Luo, J.~w.~Mei and Z.~z.~Xing,
  ``Radiative generation of leptonic CP violation,''
  Phys.\ Rev.\ D {\bf 72}, 053014 (2005)
  [hep-ph/0507065].


\bibitem{Xing:2005fw}
  Z.~z.~Xing,
  ``A Novel parametrization of tau-lepton dominance and simplified one-loop renormalization-group equations of neutrino mixing angles and CP-violating phases,''
  Phys.\ Lett.\ B {\bf 633}, 550 (2006)
  [hep-ph/0510312].

\bibitem{Xing:2011zza}
  Z.~z.~Xing and S.~Zhou,
  ``Neutrinos in particle physics, astronomy and cosmology,''
  Springer-Verlag, Berlin Heidelberg (2011).

\bibitem{Ohlsson:2012pg}
  T.~Ohlsson, H.~Zhang and S.~Zhou,
  ``Radiative corrections to the leptonic Dirac $CP$-violating phase,''
  Phys.\ Rev.\ D {\bf 87}, no. 1, 013012 (2013)
  [arXiv:1211.3153].

\bibitem{Ohlsson:2013xva}
  T.~Ohlsson and S.~Zhou,
  ``Renormalization group running of neutrino parameters,''
  Nature Commun.\  {\bf 5}, 5153 (2014)
  [arXiv:1311.3846].

\bibitem{Xing:2017mkx}
Z.~z.~Xing, D.~Zhang and J.~y.~Zhu,
``The $\mu-\tau$ reflection symmetry of Dirac neutrinos and its breaking effect via quantum corrections,''
JHEP \textbf{11}, 135 (2017)
[arXiv:1708.09144].

\bibitem{Xing:2018kto}
Z.~z.~Xing and S.~Zhou,
``Naumov- and Toshev-like relations in the renormalization-group evolution of quarks and Dirac neutrinos,''
Chin. Phys. C \textbf{42}, no.10, 103105 (2018)
[arXiv:1804.01925].

\bibitem{Zhu:2018dvj}
J.~Y.~Zhu,
``Leptonic unitarity triangles: RGE running effects and $\mu$-$\tau$ reflection symmetry breaking,''
Phys. Rev. D \textbf{99}, no.3, 033003 (2019)
[arXiv:1810.04426].

\bibitem{Zhang:2020lsd}
  D.~Zhang,
  ``Integral solutions to the one-loop renormalization-group equations for lepton flavor mixing parameters and the Jarlskog invariant,''
  arXiv:2007.12976.

\bibitem{Esteban:2020cvm}
  I.~Esteban, M.~C.~Gonzalez-Garcia, M.~Maltoni, T.~Schwetz and A.~Zhou,
  ``The fate of hints: updated global analysis of three-flavor neutrino oscillations,''
  arXiv:2007.14792.

\bibitem{Capozzi:2018ubv}
  F.~Capozzi, E.~Lisi, A.~Marrone and A.~Palazzo,
  ``Current unknowns in the three neutrino framework,''
  Prog.\ Part.\ Nucl.\ Phys.\  {\bf 102}, 48 (2018)
  [arXiv:1804.09678].

\bibitem{Endoh:2000hc}
  T.~Endoh, T.~Morozumi, T.~Onogi and A.~Purwanto,
  ``CP violation in seesaw model,''
  Phys.\ Rev.\ D {\bf 64}, 013006 (2001)
  Erratum: [Phys.\ Rev.\ D {\bf 64}, 059904 (2001)]
  [hep-ph/0012345].


\bibitem{GonzalezFelipe:2003fi}
  R.~Gonzalez Felipe, F.~R.~Joaquim and B.~M.~Nobre,
  ``Radiatively induced leptogenesis in a minimal seesaw model,''
  Phys.\ Rev.\ D {\bf 70}, 085009 (2004)
  [hep-ph/0311029].


\bibitem{Turzynski:2004xy}
  K.~Turzynski,
  ``Degenerate minimal seesaw and leptogenesis,''
  Phys.\ Lett.\ B {\bf 589}, 135 (2004)
  [hep-ph/0401219].


\bibitem{Joaquim:2005zv}
  F.~R.~Joaquim,
  ``Radiative leptogenesis in minimal seesaw models,''
  Nucl.\ Phys.\ Proc.\ Suppl.\  {\bf 145}, 276 (2005)
  [hep-ph/0501221].

\bibitem{Branco:2005ye}
  G.~C.~Branco, R.~Gonzalez Felipe, F.~R.~Joaquim and B.~M.~Nobre,
  ``Enlarging the window for radiative leptogenesis,''
  Phys.\ Lett.\ B {\bf 633}, 336 (2006)
  [hep-ph/0507092].


\bibitem{Dev:2014laa}
  P.~S.~Bhupal Dev, P.~Millington, A.~Pilaftsis and D.~Teresi,
  ``Flavour Covariant Transport Equations: an Application to Resonant Leptogenesis,''
  Nucl.\ Phys.\ B {\bf 886}, 569 (2014)
  [arXiv:1404.1003].


\bibitem{Pilaftsis:2015bja}
  A.~Pilaftsis and D.~Teresi,
  ``Mass bounds on light and heavy neutrinos from radiative minimal-flavor-violation leptogenesis,''
  Phys.\ Rev.\ D {\bf 92}, no. 8, 085016 (2015)
  [arXiv:1506.08124].

\bibitem{Pilaftsis:2003gt}
  A.~Pilaftsis and T.~E.~J.~Underwood,
  ``Resonant leptogenesis,''
  Nucl.\ Phys.\ B {\bf 692}, 303 (2004)
  [hep-ph/0309342].

\bibitem{Pilaftsis:2005rv}
  A.~Pilaftsis and T.~E.~J.~Underwood,
  ``Electroweak-scale resonant leptogenesis,''
  Phys.\ Rev.\ D {\bf 72}, 113001 (2005)
  [hep-ph/0506107].

\bibitem{Anisimov:2005hr}
A.~Anisimov, A.~Broncano and M.~Plumacher,
``The CP-asymmetry in resonant leptogenesis,''
Nucl. Phys. B \textbf{737}, 176-189 (2006)
[arXiv:hep-ph/0511248].


\bibitem{Dev:2017wwc}
B.~Dev, M.~Garny, J.~Klaric, P.~Millington and D.~Teresi,
Int. J. Mod. Phys. A \textbf{33}, 1842003 (2018)
[arXiv:1711.02863].

\bibitem{Zhang:2015tea}
  J.~Zhang and S.~Zhou,
  ``A Further Study of the Frampton-Glashow-Yanagida Model for Neutrino Masses, Flavor Mixing and Baryon Number Asymmetry,''
  JHEP {\bf 1509}, 065 (2015)
  [arXiv:1505.04858].

\bibitem{Kleppe:1995zz}
  A.~Kleppe,
  ``Extending the standard model with two right-handed neutrinos,''
in {\it Neutrino physics. Proceedings of 3rd Tallinn Symposium}, Lohusalu, Estonia, October 8-11, 1995, page 118-125.

\bibitem{Ma:1998zg}
E.~Ma, D.~P.~Roy and U.~Sarkar,
``A Seesaw model for atmospheric and solar neutrino oscillations,''
Phys. Lett. B \textbf{444}, 391-396 (1998)
[arXiv:hep-ph/9810309].

\bibitem{King:1999mb}
S.~F.~King,
``Large mixing angle MSW and atmospheric neutrinos from single right-handed neutrino dominance and U(1) family symmetry,''
Nucl. Phys. B \textbf{576}, 85-105 (2000)
[arXiv:hep-ph/9912492].

\bibitem{King:2002nf}
S.~F.~King,
``Constructing the large mixing angle MNS matrix in seesaw models with right-handed neutrino dominance,''
JHEP \textbf{09}, 011 (2002)
[arXiv:hep-ph/0204360].

\bibitem{Frampton:2002qc}
P.~H.~Frampton, S.~L.~Glashow and T.~Yanagida,
``Cosmological sign of neutrino CP violation,''
Phys. Lett. B \textbf{548}, 119-121 (2002)
[arXiv:hep-ph/0208157].

\bibitem{Guo:2006qa}
W.~l.~Guo, Z.~z.~Xing and S.~Zhou,
``Neutrino Masses, Lepton Flavor Mixing and Leptogenesis in the Minimal Seesaw Model,''
Int. J. Mod. Phys. E \textbf{16}, 1-50 (2007)
[arXiv:hep-ph/0612033].

\bibitem{Xing:2020ald}
Z.~z.~Xing and Z.~h.~Zhao,
``The minimal seesaw and leptogenesis models,''
[arXiv:2008.12090].

\bibitem{Casas:2001sr}
J.~A.~Casas and A.~Ibarra,
``Oscillating neutrinos and $\mu \to e \gamma$,''
Nucl. Phys. B \textbf{618}, 171-204 (2001)
[arXiv:hep-ph/0103065].

\bibitem{Ibarra:2003up}
A.~Ibarra and G.~G.~Ross,
``Neutrino phenomenology: The Case of two right-handed neutrinos,''
Phys. Lett. B \textbf{591}, 285-296 (2004)
[arXiv:hep-ph/0312138].




\end{thebibliography}
\end{document}